\documentclass[nofootinbib,aps]{revtex4}

\usepackage{amssymb,amstext,amsmath,amsthm}
\usepackage[dvips]{graphicx}
\usepackage{amsfonts}
\usepackage{bbm}
\usepackage{axodraw}
\usepackage{color}

\newcommand{\Ref}[1]{(\ref{#1})}

\newcommand{\eqa}{\begin{eqnarray}}
\newcommand{\neqa}{\end{eqnarray}}
\newcommand{\equ}{\begin{equation}}
\newcommand{\nequ}{\end{equation}}
\newcommand{\no}{\nonumber\\}

\def\om{\omega}
\def\w{\wedge}
\def\la{\langle}
\def\ra{\rangle}

\newcommand{\mean}[1]{\la{#1}\ra}
\newcommand{\6}{$\{6j\}$}

\newcommand{\p}{\partial}

\def\d{\delta}
\def\f{\frac}

\usepackage{bbm}

\newcommand{\scr}{\rm\scriptscriptstyle}

\newcommand{\lalg}[1]{\mathfrak{#1}}  
\newcommand{\SU}{\mathrm{SU}}

\newcommand{\su}{\lalg{su}}

\let\eps=\epsilon
\newcommand{\lp}{\ell_{\rm P}}

\newcommand{\sorgenti}{
\SetScale{0.1}\SetWidth{1.5}
\SetColor{Black}\Line(211,376)(105,46)\Line(105,46)(360,-45)\Line(360,-45)(211,376)\Line(211,376)(479,122)
\Line(479,122)(360,-44)\DashLine(105,46)(479,122){10}\SetWidth{0.5}\GOval(350,242)(37,37)(0){0.882}
\GOval(285,168)(37,37)(0){0.882}\GOval(235,-7)(37,37)(0){0.882}\GOval(429,48)(37,37)(0){0.882}
\GOval(271,79)(37,37)(0){0.882}\GOval(158,207)(37,37)(0){0.882}
\Text(12,18)[lb]{\tiny{\Black{$J_3$}}}\Text(20,-3)[lb]{\tiny{\Black{$J_1$}}}
\Text(23,5)[lb]{\tiny{\Black{$J_5$}}}\Text(39,2)[lb]{\tiny{\Black{$J_6$}}}
\Text(24,14)[lb]{\tiny{\Black{$J_2$}}}\Text(31,22)[lb]{\tiny{\Black{$J_4$}}}
}

\newcommand{\tetnet}{
\SetWidth{3}\SetScale{0.15}\SetColor{Black}\CArc(189,79)(62.97,43,403)\Line(189,141)(189,216)\Line(189,216)(346,216)
\Line(346,216)(346,-60)\Line(346,-60)(189,-60)\Line(189,-60)(189,16)\Line(126,80)(252,80)
\Text(21,25)[lb]{\tiny{{$j_1$}}}\Text(37,18)[lb]{\tiny{{$j_2$}}}\Text(13,18)[lb]{\tiny{{$j_3$}}}
\Text(24,5)[lb]{\tiny{{$j_4$}}}\Text(37,2)[lb]{\tiny{{$j_6$}}}\Text(13,2)[lb]{\tiny{{$j_5$}}}
}

\begin{document}

\title{\Large\bf Grasping rules and semiclassical limit of the geometry \\ in the Ponzano--Regge model}
\author{{\bf Jonathan Hackett and Simone Speziale}\footnote{jhackett@perimeterinstitute.ca, sspeziale@perimeterinstitute.ca}}
\affiliation{Perimeter Institute, 31 Caroline St. N, Waterloo, ON N2L 2Y5,
Canada.}

\begin{abstract}{
\noindent {We show how the expectation values of geometrical
quantities in 3d quantum gravity can be explicitly computed using
grasping rules. We compute the volume of a labelled
tetrahedron using the triple grasping. We show that the large spin
expansion of this value is dominated by the classical expression,
and we study the next to leading order quantum corrections. }}
\end{abstract}

\maketitle



\section{Introduction}
The spinfoam non--perturbative approach to quantum gravity
\cite{carlo} describes a microscopical quantum geometry, where the
geometrical observables have discrete values, expressed in terms of
half--integers (spins). These spins characterize ``atoms'' of spacetime.
A possible way to study the semiclassical limit consists in studying the
expansion for large values of the spins. In this paper, we apply this procedure to
the volume of a tetrahedron in the Ponzano--Regge (PR) model for 3d
Riemannian quantum gravity \cite{Ponzano}.

The spins, which are the fundamental variables of the model, label
the irreducible representations (irreps) of $\SU(2)$, the gauge
group of 3d GR. Using the generating functional introduced in
\cite{FreidelSFM}, the geometrical expectation values can be defined
via the action of grasping operators. This action can be evaluated
using the recoupling theory of $\SU(2)$, and thus the geometrical
values given in terms of purely algebraic quantities. Here we show
that the relevant graspings can be identified starting from the
discretization of the classical observables, and we focus on the
first non--trivial observable, the quantum volume of a tetrahedron.
Its value is obtained by triple graspings acting on the \6 symbol
associated with the tetrahedron. We identify all the relevant
graspings, and evaluate their action to write explicitly the value
of the quantum volume in terms of algebraic quantities. This is our
first result. The relevance of this result concerns the construction
of spinfoam models of matter coupled to quantum gravity. The matter
action typically contains a volume term, and thus knowing the value
of the quantum volume of a labelled tetrahedron is needed for
constructing the coupled model. For instance, this result can be
used in \cite{Fairbairn} and \cite{YM}. Furthermore, a power series
of triple graspings acting on the $\SU(2)$ \6 symbol can be used to
reconstruct the quantum \6 symbol used in the Turaev--Viro model
\cite{FK}.

Studying the large spin expansion of this value, we identify the
dominant and subdominant graspings, and show that the expansion is
remarkably dominated by the classical formula. However, the exact
result for the volume has an extra factor multiplying the classical
formula. This factor can be understood in terms of the well--known
feature of the PR model to sum over both orientations of spacetime.
The consequences of this fact for the volume were discussed in
\cite{FK}, and the factor here obtained agrees with their results.
Motivated by the analysis of \cite{FK}, we then consider the
\emph{squared} volume, and show that the correct classical formula
is this time directly reproduced. This is our second result. The
relevance of this result is to support the idea that the
semiclassical limit of spinfoam models can be studied considering
the large spin expansion. This idea is also supported by the recent
calculations of the 2-point function \cite{graviton, Io, 3dcorr}.
This is the main open problem in the formalism, and other remarkable
ideas to address it include defining coarse graining procedures
\cite{LivineTerno}, contructing effective field theories
\cite{EteraPRL}, or rewriting conventional quantum field theories in
the language of spinfoams \cite{Baratin}.

This paper is organized as follows. In Section \ref{SectionPR}, we
briefly recall how the Ponzano--Regge model is constructed, and we
introduce the generating functional to compute expectation values of
the geometry. In Section \ref{section3dQG}, we show how to introduce
the geometrical observables using the discrete variables of the
model, and how to relate this observables to grasping operators. We
compute the values of quantum lengths and angles, corresponding to
quadratic graspings, and the value of the quantum volume,
corresponding to the triple grasping. In Section \ref{semi3d}, we
analyze the leading order of the large spin expansion for the volume
and the squared volume. In Section \ref{SectionCorr}, we compute the
next to leading order quantum corrections. In the final Section
\ref{SectionConcl} we summarize our results. All the calculations
used in the paper are explicitly reported in the Appendix, where we
also fix our phase convention for the evaluation of spin networks,
and show how this is related to the grasping rules.

Throughout we define the Planck length as $\ell_{\rm P}:=16 \pi
\hbar G$.

\section{Ponzano--Regge model and generating functional}\label{SectionPR}

We consider here the PR model on a fixed triangulation $\Delta$ of
the spacetime manifold $M$. This can be obtained from a path
integral quantization of the first order triad action for Riemannian
GR. In the continuum, the fundamental fields are the $\SU(2)$
connection $\om^{IJ}_\mu$ and the triad $e_\mu^I$, and the action
for GR reads \equ\label{actionGR} S_{\rm GR}[e_\mu^I(x),
\omega_{\mu}^{IJ}(x)] =\frac{1}{16\pi G} \int_M \;{\rm Tr} \;e\wedge
F(\omega), \nequ where the trace Tr is over the $\SU(2)$ indices,
and $F(\om)$ is the curvature of $\om$. This action can be
discretised on $\Delta$ as follows. Consider an embedding $i:\Delta
\rightarrow M$, which allows to think of $\Delta$ as a cellular
decomposition of $M$. Using this embedding, we can define the
vectors $\ell_s^\mu \sim \int_{s} dx^\mu$, tangent to the segments
$s$ of $\Delta$. Analogously, we can define the vectors $\ell_e^\mu
\sim \int_{e} dx^\mu$ tangent to the edges $e$ of the dual
triangulation $\Delta^*$ (which we recall are in one to one
correspondence with triangles of $\Delta$).

We then discretise the triad field with an ${\su}(2)$ algebra element, associated to the segments of $\Delta$,
and the connection with an $\SU(2)$ element, associated to the edges of $\Delta^*$:
\eqa\label{defX}
e_\mu^I(x) \mapsto X_s^I &:=&
\frac{1}{\ell_{\rm P}} e_\mu^I(x)\ell_s^\mu \sim \frac{1}{\ell_{\rm P}} \int_s\,e_\mu^I(x) dx^\mu.
\\
\label{defg} \omega_\mu^{IJ}(x)\mapsto g_e &:=&
e^{\omega_\mu^{IJ}\ell_{e}^\mu} \sim e^{\int_{e} \omega^{IJ}}. \neqa
We also introduce the quantities $g_f = \prod_{e\in\p f}g_e$
associated with the faces $f$ of $\Delta^*$. We see from \Ref{defg}
that these quantities discretise the curvature, $g_f\sim \exp \int_f
F(\om)$. Recalling that faces of $\Delta^*$ are in one to one
correspondence with segments of $\Delta$, we will use from now on
the notation $g_s\equiv g_f$.

Consequently, we write the discrete action as
\equ\label{discreteBF}
S[X_s^I, g_{e}] = \hbar \sum_s {\rm Tr} \;[X_s \,g_s].
\nequ
When the embedding is sufficiently refined, and the coordinate areas consequently small,
we can expand $g_s\sim {\mathbbm 1} + \int_f F(\om)$. Recalling that Tr $T=0$ for any $T\in\su(2)$,
we see that \Ref{discreteBF} reduces to \Ref{actionGR}.

The quantum theory can be constructed from the partition function
\equ\label{Z} Z = \prod_{e} \int_{\SU(2)} dg_{e} \; \prod_s
\int_{{\mathfrak su}(2)}dX_s \; e^{{i} \sum_s {\rm Tr} \, [X_s
\,g_s]}. \nequ This quantity can be evaluated using the harmonic
analysis of $\SU(2)$ (for details, see for instance \cite{carlo}),
and one obtains \equ\label{ZPR} Z = \sum_{j_s} \prod_s d_{j_s}
\prod_\tau \{6j\}, \nequ where the sum is over all possible
assignments of half--integers $j$ to the segments of $\Delta$. The
half--integers, or spins, label the irreducible representations of
$\SU(2)$. The quantity $d_j:=2j+1$ is the dimension of the
representation. Finally, a \6 symbol is associated with each
tetrahedron $\tau$ of $\Delta$. For more discussion of the PR model,
see \cite{Laurent3}. The \6 symbol is the key object of the
recoupling theory of $\SU(2)$, and it depends only on the six $j$s
attached to the segments of the tetrahedron. It can be written in
terms of the Wigner 3m--symbols as \equ\label{combin6j}
\left\{\begin{array}{ccc} j_1 & j_2 & j_3 \\
j_4 & j_5 & j_6 \end{array} \right\} :=
\left( \begin{array}{ccc} j_1 & j_2 & j_3 \\
m_1 & m_2 & m_3 \end{array} \right)
\left( \begin{array}{ccc} j_1 & j_5 & j_6 \\
m_1 & m_5 & m_6 \end{array} \right)
\left( \begin{array}{ccc} j_4 & j_5 & j_3 \\
m_4 & m_5 & m_3 \end{array} \right)
\left( \begin{array}{ccc} j_4 & j_2 & j_6 \\
m_4 & m_2 & m_6 \end{array} \right).
\nequ

The \6 symbol has a well--known asymptotic behaviour, namely that if
we rescale the half--integers entering the $\{6j\}$ symbol as $j_s
\equiv N k_s$, we have \cite{Ponzano, asympt, asympt2}
\equ\label{asymp} \lim_{N\mapsto\infty}\{6j\} \;=\; \f1{\sqrt{12\pi
V(j_s)}} \cos\left(\sum_s (j_s+\frac{1}{2})\theta_s(j_s) +
\frac{\pi}{4} \right), \nequ where $V(j_s)$ is the classical volume
of the tetrahedron with segment lengths given by $j+\f12$, and
$\theta_s$ are the corresponding dihedral angles, namely the angles
between the normals to the triangles. As it will be useful in the
following, we recall here that the volume of a tetrahedron can be
expressed in terms of a dihedral angle using the formula
\begin{equation}\label{V}
V = \f23\f{A_1 A_2}{\ell_s} \sin\theta_s,
\hspace{2cm} \parbox[3cm]{4cm}{\includegraphics[width=3cm]{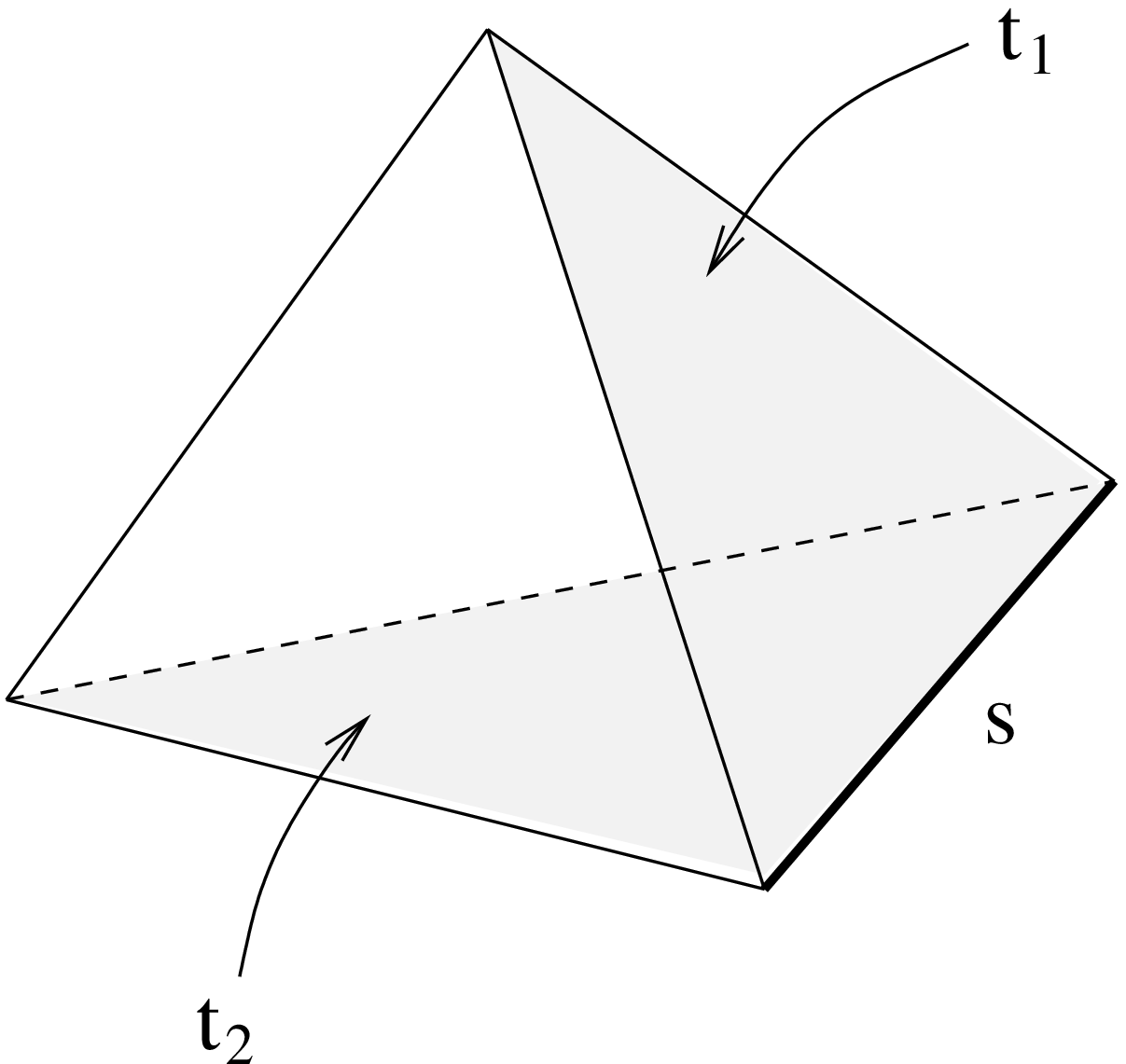}}
\end{equation}
where $A_1$ and $A_2$ are the areas of the two triangles $t_1$ and $t_2$ sharing
the segment $s$, as shown in the above figure.
The areas can be expressed in terms of the segment lengths, using Heron's formula
(see the Appendix).

The key point of \Ref{asymp} is the argument of the cosine:
up to the factor $\f\pi4$ (which does not change the equations of motion), this is the Regge action
$S_{\rm R}=\sum_s \ell_s \theta_s(\ell_s)$, a discrete approximation to classical GR.
Therefore, the amplitude of $Z$ is dominated by exponentials of the Regge action,
which makes it promising to study the semiclassical limit in this way.
If this is correct, then also the geometrical
quantities that one can evaluate in the model should reduce to their classical expressions, in
the large $j$ limit defined above.

However one would expect a single exponential to arise in the semiclassical limit, 
namely $Z_{\scr GR}\sim \int {\cal D}g\, e^{iS_{\rm R}}$.
The meaning of this difference is well studied, see for instance the discussion in \cite{carlo}.
We might then conclude that on a single tetrahedron $\tau$ $Z(\tau)=Z_{\scr GR}(\tau)+\overline{Z_{\scr GR}(\tau)}$.
As we will see below, this fact has consequences for the expectation value of the volume.

To introduce the geometrical observables in the PR model,
below we write them as gauge--invariant functions of the variables $X_s^I$.
Then, to compute their expectation values, it is convenient to introduce a generating
functional,
\equ\label{genBF}
Z[J] = \prod_{e} \int_{\SU(2)} dg_{e} \; \prod_s \int_{{\mathfrak su}(2)}dX_s
\; e^{{i} \sum_s {\rm Tr} \, X_s \left(g_s + J_s \right)}.
\nequ
This can be evaluated as described in \cite{FreidelSFM}, to give
\vspace{0.7cm}
\equ\label{gen}
Z[J] = \sum_{\{j_s\}}\; \prod_s d_{j_s} \;\prod_\tau
\parbox[2cm]{1.8cm}{\fcolorbox{white}{white}{
\begin{picture}(0,0) (10,15) \sorgenti \end{picture}}},
\nequ
\vspace{0.3cm}
where
\vspace{0.5cm}
\eqa\label{6jsorgenti}
\parbox[2cm]{1.8cm}{\fcolorbox{white}{white}{
\begin{picture}(0,0) (10,15) \sorgenti \end{picture}}} &:=&
\left( \begin{array}{ccc} j_1 & j_2 & j_3 \\
m_1 & m_2 & m_3 \end{array} \right)
\left( \begin{array}{ccc} j_1 & j_5 & j_6 \\
n_1 & m_5 & m_6 \end{array} \right)
\left( \begin{array}{ccc} j_4 & j_5 & j_3 \\
m_4 & n_5 & n_3 \end{array} \right)
\left( \begin{array}{ccc} j_4 & j_2 & j_6 \\
n_4 & n_2 & n_6 \end{array} \right) \times \no\no &&
D^{(j_1)}_{m_1n_1}(e^{J_1}) D^{(j_2)}_{m_2n_2}(e^{J_2})
D^{(j_3)}_{m_3n_3}(e^{J_3}) D^{(j_4)}_{m_4n_4}(e^{J_4})
D^{(j_5)}_{m_5n_5}(e^{J_5}) D^{(j_6)}_{m_6n_6}(e^{J_6}). \neqa Here
the $D$ are representation matrices. The quantity \Ref{6jsorgenti}
represents the \6 symbols with source insertions. The $J$'s are
attached to the segments of $\Delta$; they are the sources of the
quantum excitations $j_s$. For all $J_s=0$ \Ref{6jsorgenti} reduces
to the expression for the $\{6j\}$ symbol given above, thus $ Z_{\rm
BF}=Z[J]\Big|_{J=0}.$

Consider now a gauge--invariant observable constructed from the $X^I_s$ variables,
$\Phi[X_s^I]$. Using the generating functional, its expectation value can be written as
\equ
\mean{\Phi} = \prod_{e} \int_{\SU(2)} dg_{e} \; \prod_s \int_{{\mathfrak su}(2)}dX_s
\; \Phi[X_s^I] \; e^{{i} \sum_s X_s^I g_s^I} =
\Phi[-i\f{\d}{\d J_s^I}] Z[J]\Big|_{J=0}.
\nequ
In the next Section, we will use this procedure to compute expectation values.
To do so, we need to know the action
of the algebra derivatives. Acting on a group element in the representation $j$, we have
\equ\label{grasp}
\f\d{\d J^I} D^{(j)}(e^{J})\Big|_{J=0} = -iT^{I(j)},
\nequ
where $T^{I(j)}$ is the $I$-th generator in the representation $j$. By inspecting \Ref{6jsorgenti},
we see that a derivative acting on $J_s$ attaches to
the segment $s$ an algebra generator in the irrep $j_s$ labeling the segment.
This action is called ``grasping'', and it is described in more details in the Appendix.
In particular, we are interested in quadratic and cubic gauge--invariant functions $\Phi$, such as squared lengths
and volumes, which are related to the action of quadratic and cubic graspings.

\section{Graspings and quantum geometry}\label{section3dQG}

\subsection{Quadratic graspings: lengths and angles}
\label{Sectiondouble}
The classical geometrical observables can be described using the discrete variables $X_s^I$.
For instance, the length of a segment $\tilde s$ can be written as
$\ell_{\tilde s}^2 = g_{\mu\nu}\ell^\mu_{\tilde s} \ell^\nu_{\tilde s} = \ell_{\rm P}^2 X_{\tilde s}^IX_{\tilde s}^I$.
In the same way, we can study the angles between the segments. To this aim, we consider
the two segments $s_1$ and $s_2$, sharing a vertex, and the segment
$s_3$ closing the triangle. The angle can be read from the scalar
product $\ell_{s_1}\cdot \ell_{s_2} = \ell_{\rm P}^2 X^I_{s_1} X^I_{s_2}$.

Calculating the expectation value of quadratic functions of the $X_s^I$s is particularly
simple. The expectation value of a scalar product is given by
\eqa\label{meanl}
\mean{\ell_{s_1}\cdot\ell_{s_2}} = \frac{1}{Z} \ell_{\rm P}^{2}
\prod_{e} \int_{\SU(2)} dg_{e} \; \prod_s \int_{{\mathfrak su}(2)}dX_s
\; X_{s_1}^I X_{s_2}^I \;e^{i \sum_s X_s^I g_s^I} =
- \frac{1}{Z} \ell_{\rm P}^2\;
\frac{\delta}{\delta J_{s_1}^I} \frac{\delta}{\delta J_{s_2}^I} \;Z[J]\;\Big|_{J=0}.
\neqa
To evaluate this quantity, we need the action of the double grasping, which is computed in the Appendix.

For the case $s_1=s_2\equiv\tilde s$, the relevant grasping gives

\equ\label{GraspingCas}
\frac{\delta}{\delta J_{\tilde s}^I} \frac{\delta}{\delta J_{\tilde s}^I}
\parbox[2cm]{1.8cm}{\begin{picture}(0,0) (10,15) \sorgenti \end{picture}} = -C^2(j_{\tilde s}) \, \{6j\}.
\nequ
Consequently, we have
\equ
\mean{\ell_{\tilde s}^2} = \f1Z \sum_{\{j_s\}}\; \lp^2 C^2(j_{\tilde s}) \
\prod_s d_{j_s} \;\prod_\tau  \,\{6j\}
\nequ
We see that the value of the quantum squared length of a labelled segment
is given by $\ell_{\rm P}^2 C^2(j_{\tilde s})$.
This is the basic result of quantum geometry, and we wee that the origin of its discreteness
lies in the compactness of the group $\SU(2)$, which gives a discrete series
of representations. Therefore, $\ell_s$ acquires only discrete values, consistently with the canonical
result \cite{carlo}.

For the case when $s_1\neq s_2$ share a vertex, the relevant grasping gives

\equ\label{Grasping2}
\frac{\delta}{\delta J_{s_1}^I} \frac{\delta}{\delta J_{s_2}^I}
\parbox[2cm]{1.8cm}{\begin{picture}(0,0) (10,15) \sorgenti \end{picture}} =
-\f{1}{2} [C^2(j_{s_1}) +C^2(j_{s_2}) -C^2(j_{s_3})] \, \{6j\}_\tau,
\nequ where $s_3$ closes the triangle with $s_1$ and $s_2$.
Therefore, we see that the value of $\cos\theta_{s_1s_2} $ is given
by the expression \equ\label{angle} \f{1}{2C(j_{s_1})C(j_{s_2})}
[C^2(j_{s_1}) +C^2(j_{s_2})- C^2(j_{s_3})]. \nequ Notice that this
expression coincides with the classical one, once we identify the
Casimir with the segment lengths. However, this does not mean that
the angle behaves classically: as the Casimir is discrete, the angle
cannot range continously between $0$ and $\pi$, instead only
specific values are allowed. Among these, the equilateral case: if
we set all $j_s=j_0$ in \Ref{angle}, we obtain $\f{1}{2} =
\cos(\f{\pi}{3})$, which correctly reproduces an equilateral
triangle.

\subsection{Triple grasping: volume}\label{Sectiontriple}
We now come to the explicit computation of the volume, which is the first non trivial geometrical
object in the PR model.
This is defined as
\equ\label{v}
{\cal V}_\tau:= \int_\tau e d^3x\simeq \f1{3!}e \,\int_\tau d^3x,
\qquad e=\f1{3!}\eps^{\mu\nu\rho}\eps_{IJK} e_\mu^I e_\nu^J e_\rho^K.
\nequ
The reason for the factor $3!$ lies in the fact the the determinant is the infinitesimal
(metrical) volume of a cube, and there are $6$ tetrahedra in a cube.
To express ${\cal V}_\tau$ in terms of the variables $X_s^I$, we proceed as follows.
Consider three segments sharing a point $p$ of $\tau$, with coordinate
vectors $\ell_s^\mu(p)$; the coordinate volume is $\int_\tau d^3x={\rm det}\, \ell_s^\mu(p)$,
by definition of $\ell_s^\mu(p)$, independently of the point considered.
We can think of $\ell^\mu_s(p)$ as a 3 by 3 matrix, with inverse $n^s_\mu(p)$ being defined by
\equ\label{unity1}
\sum_{s\in p(\tau)}\ell_s^\mu(p) n_\nu^{s}(p) = \delta^\mu_\nu.
\nequ
This resolution of the identity can now be inserted in the definition \Ref{v} of $e$, in order
to give the variables $X_s^I$:
$$
{\cal V}_\tau = \f{1}{3!} \, e\, \det \ell^\mu_s =
\f{\det \ell^\mu_s}{3!} \frac{\ell_{\rm P}^3}{3!}
\sum_{s_1\in p_1} \sum_{s_2\in p_2} \sum_{s_3\in p_3} \eps_{IJK} \, \eps^{\mu\nu\rho}\,
n_\mu^{s_1} n_\nu^{s_2} n_\rho^{s_3} \, X_{s_1}^I X_{s_2}^J X_{s_3}^K.
$$
Notice that from the definition of $n_\mu^s(p)$ in \Ref{unity1}, we have
$\epsilon^{\mu\nu\rho}n^{s_1}_\mu n^{s_2}_\nu n^{s_3}_\rho=
\epsilon^{s_1s_2s_3}({\rm det}\, \ell_s^\mu)^{-1}$, thus
$$
{\cal V}_\tau = \frac{\ell_{\rm P}^3}{3!^2}
\sum_{s_1\in p_1} \sum_{s_2\in p_2} \sum_{s_3\in p_3} \eps_{IJK} \, \eps^{s_1s_2s_3}\,
X_{s_1}^I X_{s_2}^J X_{s_3}^K.
$$
With an eye at the construction of its quantum version, it is
convenient to symmetrise this expression. One possible way to do so
is to take the same point $p$ in each insertion of \Ref{unity1}, and
then sum over the four contributions: \equ\label{volume} {\cal
V}_\tau = \frac{\ell_{\rm P}^3}{3!^2} \f14\sum_{p\in\tau}
\sum_{s_1,s_2,s_3\in p}\eps_{IJK} \, \eps^{s_1s_2s_3}\, X_{s_1}^I
X_{s_2}^J X_{s_3}^K =\f{\lp^3}{3!}\f14 \sum_{p\in\tau}\eps_{IJK} \,
X_{s_1}^I X_{s_2}^J X_{s_3}^K, \nequ where in the latter expression
$s_1$, $s_2$ and $s_3$ are a fixed right--handed triple belonging to
the given $p$.

Alternatively, we can consider all possible sixteen non--coplanar triples of segments, and write
\equ\label{volume1}
{\cal V}_\tau =\f1{3!}\f{\lp^3}{16} \sum_{s_1,s_2,s_3}\eps_{IJK} \, X_{s_1}^I X_{s_2}^J X_{s_3}^K.
\nequ
This formula was proposed in \cite{FreidelSFM}.
Notice that this latter case is more generic:
\Ref{volume} can be obtained from \Ref{volume1} restricting the triplets in the sum.
The two expressions are clearly classically equivalent, but lead to different quantum values.
As we show below, the corresponding values in the quantum theory have the same semiclassical leading order, but different corrections.

As it is more generic, we consider first the expression \Ref{volume1}.
Using the generating functional as above, we have
\eqa\label{exV}
\langle {\cal V}_{\tilde\tau} \rangle &=& \frac{1}{Z}\, \f1{3!}\frac{\ell_{\rm P}^{3}}{16}
\prod_{e} \int_{\SU(2)} dg_{e} \; \prod_s \int_{{\mathfrak su}(2)}dX_s
\; \sum_{s_1,s_2,s_3} \eps_{IJK}X^I_{s_1}X^J_{s_2}X^K_{s_3} \;e^{i \sum_s X_s^I g_s^I}
= \no &=& \frac{1}{Z}\,\f i{3!}
\frac{\ell_{\rm P}^{3}}{16} \left(\sum_{s_1,s_2,s_3} \eps_{IJK}\frac{\delta}{\delta J_{s_1}^I}
\frac{\delta}{\delta J_{s_2}^J} \frac{\delta}{\delta J_{s_3}^K}\right)\;Z[J]\;\Big|_{J=0}.
\neqa
The derivatives act only on the sources
present in the tetrahedron $\tilde\tau$. Therefore, $\mean{{\cal V}_{\tilde \tau}}$
depends only on the six $j$s entering $\tilde \tau$.
The quantity in round brackets gives rise to the triple grasping on the tetrahedron,
as explained in the Appendix. Depending
on the configuration of the triplet $s_1,s_2,s_3$ considered, different types of grasping are involved.
These types, together with the result of their evaluation, are listed in Table \ref{table}.
We report the details of the evaluation in the Appendix, and discuss here the results.
First of all, notice the colour coding: we used red for those graspings, type 2 and 3, corresponding
to non coplanar triplets of different segments, which would be classically used to compute the volume.
Blue, on the other hand, is used for configurations which classically would not contribute.
These correspond to cases when the three segments are coplanar (type 1), or when two
(type 4 and 5) and even all three (type 6) are the same.
As in the classical case, coplanar triplets
do not contribute to the volume (and this is the reason why we did not
include this grasping in \Ref{volume1}).
On the contrary, the classically absent types 4, 5 and 6 do contribute to the volume,
but only at the level of quantum corrections, as they scale as $j^2$ (see column on the
right of the Table). Indeed,
the relevant graspings for the semiclassical limit are only the types 2 and 3, as they scale as $j^3$.

We see from the results listed in the table that evaluating the action of the triple grasping is more involved
than the quadratic one. In particular, notice that the graspings 2, 3 and 5 do not preserve the original tetrahedral amplitude \6, but contain a superposition of terms.

\begin{table}[t]
\framebox{\begin{tabular}{cc|c|c}
&{\bf Grasping} &  {\bf Evaluation} & {\bf Leading order} \\
& \hspace{2cm} & \hspace{11cm} & \hspace{3cm}  \\ & & & \\ 1. &
\parbox[2cm]{2cm}{\begin{picture}(0,0)(30,15)\SetScale{0.1}\SetWidth{1.5}
\SetColor{Black}\Line(211,376)(105,46)\Line(105,46)(360,-45)\Line(360,-45)(211,376)\Line(211,376)(479,122)
\Line(479,122)(360,-44)\DashLine(105,46)(479,122){10}
\SetWidth{4}\SetColor{Blue}\Line(220,120)(230,1)\Line(220,120)(293,138)\Line(220,120)(150,180)
\end{picture}} & 0 & 0 \\ & & & \\ & & & \\ & & & \\ 2. &
\parbox[2cm]{2cm}{\begin{picture}(0,0)(30,15)\SetScale{0.1}\SetWidth{1.5}
\SetColor{Black}\Line(211,376)(105,46)\Line(105,46)(360,-45)\Line(360,-45)(211,376)\Line(211,376)(479,122)
\Line(479,122)(360,-44)\DashLine(105,46)(479,122){10}
\SetWidth{4}\SetColor{Red}\Line(220,120)(230,1)\Line(220,120)(293,138)\Line(220,120)(404,20)
\end{picture}} & \small{$c_-(j_s) \left\{\begin{array}{ccc} j_1-1 & j_2 & j_3 \\
j_4 & j_5 & j_6 \end{array} \right\} + c_0(j_s) \left\{\begin{array}{ccc} j_1 & j_2 & j_3 \\
j_4 & j_5 & j_6 \end{array} \right\} + c_+(j_s) \left\{\begin{array}{ccc} j_1+1 & j_2 & j_3 \\
j_4 & j_5 & j_6 \end{array} \right\}$}
& $j^3$ \\ & & & \\ & & & \\ & & & \\ 3. &
\parbox[2cm]{2cm}{\begin{picture}(0,0)(30,15)\SetScale{0.1}\SetWidth{1.5}
\SetColor{Black}\Line(211,376)(105,46)\Line(105,46)(360,-45)\Line(360,-45)(211,376)\Line(211,376)(479,122)
\Line(479,122)(360,-44)\DashLine(105,46)(479,122){10}
\SetWidth{4}\SetColor{Red}\Line(220,120)(230,1)\Line(220,120)(293,138)\Line(220,120)(320,270)
\end{picture}} & \small{$c_-(j_s) \left\{\begin{array}{ccc} j_1-1 & j_2 & j_3 \\
j_4 & j_5 & j_6 \end{array} \right\} + c_0(j_s) \left\{\begin{array}{ccc} j_1 & j_2 & j_3 \\
j_4 & j_5 & j_6 \end{array} \right\} + c_+(j_s) \left\{\begin{array}{ccc} j_1+1 & j_2 & j_3 \\
j_4 & j_5 & j_6 \end{array} \right\}$}
& $j^3$  \\ & & & \\ & & & \\ & & & \\ 4. &
\parbox[2cm]{2cm}{\begin{picture}(0,0)(30,15)\SetScale{0.1}\SetWidth{1.5}
\SetColor{Black}\Line(211,376)(105,46)\Line(105,46)(360,-45)\Line(360,-45)(211,376)\Line(211,376)(479,122)
\Line(479,122)(360,-44)\DashLine(105,46)(479,122){10}
\SetWidth{4}\SetColor{Blue}\Line(220,120)(230,1)\Line(220,120)(293,138)\Line(220,120)(180,20)
\end{picture}} & \small{$-\f12\left[ C^2(j_1)+C^2(j_2)-C^2(j_3) \right] \,
\left\{\begin{array}{ccc} j_1 & j_2 & j_3 \\ j_4 & j_5 & j_6  \\ \end{array}\right\}$}
& $j^2$ \\ & & & \\ & & & \\ & & & \\ 5. &
\parbox[2cm]{2cm}{\begin{picture}(0,0)(30,15)\SetScale{0.1}\SetWidth{1.5}
\SetColor{Black}\Line(211,376)(105,46)\Line(105,46)(360,-45)\Line(360,-45)(211,376)\Line(211,376)(479,122)
\Line(479,122)(360,-44)\DashLine(105,46)(479,122){10}
\SetWidth{4}\SetColor{Blue}\Line(220,120)(230,1)\Line(220,120)(370,223)\Line(220,120)(180,20)
\end{picture}} &  \small{$-\f{c_-(j_s)}{j_1+1} \left\{\begin{array}{ccc} j_1-1 & j_2 & j_3 \\ j_4 & j_5 & j_6 \end{array} \right\} + c_0(j_s) \left\{\begin{array}{ccc} j_1 & j_2 & j_3 \\ j_4 & j_5 & j_6 \end{array} \right\}
+ \f{c_+(j_s)}{j_1} \left\{\begin{array}{ccc} j_1+1 & j_2 & j_3 \\ j_4 & j_5 & j_6 \end{array} \right\}$}
& $j^2$  \\ & & & \\ & & & \\ & & & \\ 6. &
\parbox[2cm]{2cm}{\begin{picture}(0,0)(30,15)\SetScale{0.1}\SetWidth{1.5}
\SetColor{Black}\Line(211,376)(105,46)\Line(105,46)(360,-45)\Line(360,-45)(211,376)\Line(211,376)(479,122)
\Line(479,122)(360,-44)\DashLine(105,46)(479,122){10}
\SetWidth{4}\SetColor{Blue}\Line(220,120)(230,1)\Line(220,120)(140,34)\Line(220,120)(180,20)
\end{picture}} &  \small{$- C^2(j_1) \, \left\{\begin{array}{ccc} j_1 & j_2 & j_3 \\
j_4 & j_5 & j_6  \\ \end{array}\right\}$}
& $j^2$ \\ & & & \\ & & & \\ & & & \\
\end{tabular}}
\caption{The different triple graspings. The labels for the segments
are as in \Ref{gen}. In red, the classical configurations; in blue,
configurations which are absent in the classical case. The details
of the evaluations, as well as the explicit values of the
coefficients $c_-, c_0$ and $c_+$, are reported in the Appendix.
With abuse of notation, we used the same symbol $c$ for the
coefficients of types $2$, $3$ and $5$; however, a permutation of
the segment labels is involved in going from one type to the other.
For each grasping type, a single configuration is shown; the others
can be obtained by permutations of the segments. Finally, notice
that there are symmetries in the evaluation: for instance in type 2,
we chose to write the final result in terms of \6 shifted in the
variable $j_1$, but we could have just as well shifted $j_2$ or
$j_6$, and the coefficients $c_\pm(j_s)$, $c_0(j_s)$ would have
accordingly changed. \label{table}}
\end{table}

The results listed in the Table can be extended to the other configurations by permutating
the segment labels. For instance, we reported the grasping of type 2 acting on the vertex
$126$. The evaluation of the same grasping on, say, the vertex $234$ can be obtained under
the permutation $123456 \mapsto 231564$.
With this understanding, we can write
the action of \Ref{volume} as
\equ\label{spectrum1}
\f i{3!} \f{\lp^3}{4}\sum_{p\in \tau} \left[
c_-(j_s) \left\{\begin{array}{ccc} j_1-1 & j_2 & j_3 \\ j_4 & j_5 & j_6 \end{array} \right\}
+ c_0(j_s) \left\{\begin{array}{ccc} j_1 & j_2 & j_3 \\ j_4 & j_5 & j_6 \end{array} \right\}
+ c_+(j_s) \left\{\begin{array}{ccc} j_1+1 & j_2 & j_3 \\ j_4 & j_5 & j_6 \end{array} \right\}
\right] \left\{\begin{array}{ccc} j_1 & j_2 & j_3 \\ j_4 & j_5 & j_6 \end{array} \right\}^{-1},
\nequ
where the explicit values of the coefficients $c_\pm(j_s)$, $c_0(j_s)$ are given in the Appendix.
Notice that they vary when different configurations (here different points of the tetrahedron)
are considered. For
simplicity of notation we did not write explicitly the dependences of the coefficients on the configuration.

Type 2 is the only grasping entering \Ref{volume}, whereas the more generic definition \Ref{volume1}
involves all the different graspings, and the result would be ($i$ times) the sum
of all the grasping evaluations listed in the Table.
Notice that in both cases the value of the volume is purely imaginary. This was anticipated in \cite{FK}, and
can be understood as follows. Firstly, the volume is odd under change of orientation, namely
$V(\tau)=-V(-\tau)$. Secondly, the unitarity of the PR amplitude implies $Z(\tau)=\overline{Z(-\tau)}$.
Therefore $\mean{V(\tau)}=-\overline{\mean{V(\tau)}}$ which implies Re$\mean{V(\tau)}\equiv 0$.
As argued in \cite{FK}, this volume can be related to a real volume $\mean{V(\tau)}_{\scr GR}$, computed using the
$Z_{\scr GR}$ partition function defined in Section \ref{SectionPR}, which we recall satisfies
$Z= 2 \, {\rm Re} \, Z_{\scr GR}$. Formally we have
\equ
\mean{V(\tau)}Z(\tau) = \mean{V(\tau)}_{\scr GR}Z_{\scr GR}(\tau) +
\overline{\mean{V(\tau)}_{\scr GR}Z_{\scr GR}(\tau)} =
2 i \mean{V(\tau)}_{\scr GR} \, {\rm Im} \, Z_{\scr GR}(\tau),
\nequ
from which
\equ\label{Vgr}
\mean{V(\tau)} = i \, \f{{\rm Im} \, Z_{\scr GR}(\tau)}{{\rm Re} \, Z_{\scr GR}(\tau)} \, \mean{V(\tau)}_{\scr GR}
\nequ
Let us now restrict this formula to
a single tetrahedron, and assume that in the semiclassical limit
$Z_{\scr GR}(\tau)\sim \exp {i( S_{\rm R}(\tau)+\f\pi4})$. Under this assumption,
the large spin limit of \Ref{Vgr} formally reads
\equ\label{Vgr1}
\mean{V(\tau)} = i \tan\Big( S_{\rm R}(j_s)+\f\pi4 \Big) \, \mean{V(\tau)}_{\scr GR}.
\nequ
We expect the value of the volume in the PR model to obey a relation
like the one above. Also, notice that the same reasoning
applied to the \emph{squared} volume gives $\mean{V(\tau)^2}=\mean{V(\tau)^2}_{\scr GR}$,
thus the PR model should indeed reproduce the correct semiclassical limit of the squared volume.

\section{The semiclassical limit of the volume}\label{semi3d}

All the non--zero graspings (plus all possible permutations) listed in Table \ref{table} contribute
to the quantum volume. However, not all of them contribute to the leading order of the
large spin limit. Indeed, the graspings have different overall scalings, which we reported for commodity
in the column on the right. In particular we see that all the
degenerate (classically absent) graspings do not contribute to the leading
order.

Let us focus our attention to the first non--degenerate grasping, number 2 in the Table. This, we recall, is the
only grasping entering the definition \Ref{volume} of the volume. As we see from the Table, the
evaluation of the grasping gives a superposition of \6s.
To study the large spin limit, consider first the coefficients:
as we show in the Appendix, in the large spin limit we have
\eqa\label{lim1}
c_{\pm}(j_s) &\simeq& \mp \f2{l_1} A_{123} A_{156},
\\\label{lim2} c_0(j_s)  &\sim& j^2.
\neqa Here $l_1= j_1+\f12$ and $A_{123}$ is the area of the triangle
bound by the segments $l_1$, $l_2$ and $l_3$. The
asymptotics \Ref{lim1} are crucial: if we compare them with \Ref{V},
we see that we are on the right track, a factor
$\left(\sin\theta_1\right)/3$ is all that is missing to recover the
classical formula of the volume. On the other hand, \Ref{lim2} shows
that the ``diagonal'' term, the one with $j_1$ unchanged in the \6,
can be neglected at the leading order. The leading order then
emerges only from the superdiagonal and the subdiagonal, namely the
ones with $j_1\pm1$ in the \6. Interestingly, this is analogous to what
happens for the volume operator in the canonical approach of loop
quantum gravity (see for instance \cite{Brunn}).

Using \Ref{asymp} to expand the \6s entering the grasping 2, the result reads
\equ\label{vol1}
\eps_{IJK}\frac{\d}{\d J_{s_1}^I}\frac{\d}{\d J_{s_2}^J} \frac{\d}{\d J_{s_3}^K}
\parbox[2cm]{1.8cm}{\begin{picture}(0,0) (0,15) \sorgenti \end{picture}}\Bigg|_{J=0}=
\f2{l_1} \f{A_{123} A_{156}}{\sqrt{12\pi \, V(j_s)}}
\Big[\cos\Big(S_{\rm R}(j_1-1)+\f\pi4\Big) -\cos\Big(S_{\rm
R}(j_1+1)+\f\pi4\Big) \Big] + O(j^2). \nequ For large spins, the
Regge actions can be expanded. Using the well known property $\f{\p
S_{\rm R}}{\p j_s}= \theta_s$, we have \equ\label{Rexp} S_{\rm
R}(j_1\pm 1)\simeq S_{\rm R}(j_s)+\f{\p S_{\rm R}}{\p j_1}\d j_1
=S_{\rm R}(j_s)\pm \theta_1. \nequ Consequently, \equ\label{trigo}
\cos\Big(S_{\rm R}(j_1-1)+\f\pi4\Big) -\cos\Big(S_{\rm
R}(j_1+1)+\f\pi4\Big) \simeq 2 \, \sin\theta_1 \, \sin\Big(S_{\rm
R}(j_s)+\f\pi4\Big) \nequ The expansion of the Regge action has
produced the sine of the dihedral angle needed to recover the
classical formula for the volume. Putting everything together,
\Ref{vol1} reads \equ\label{vol2} \f{i}{3!}\,\eps_{IJK}\frac{\d}{\d
J_{s_1}^I}\frac{\d}{\d J_{s_2}^J} \frac{\d}{\d J_{s_3}^K}
\parbox[2cm]{1.8cm}{\begin{picture}(0,0) (0,15) \sorgenti \end{picture}}\Bigg|_{J=0}=
i \, \f{V(j_s)}{\sqrt{12\pi \, V(j_s)}}\, \sin\Big(S_{\rm R}(j_s)+\f\pi4\Big) + O(j^2).
\nequ

In the large spin limit, we then have 
\equ\label{limitV} \mean{{\cal
V}_{\tilde\tau}} \sim i \f1Z \sum_{j\gg1} \prod_s d_{j_s}
\prod_{\tau} \{6j\} \, V_{\tilde\tau}(j_s) \, \tan\Big(S_{\rm
R}(j_s)+\f\pi4\Big), \nequ 
We see that the leading order of the
quantum volume is indeed proportional to the classical formula
\Ref{V}, but there is the extra factor of the tangent of the Regge
action. This was to be expected, and it is consistent with the
formal manipulation \Ref{Vgr1}.

As discussed above, we can consider the squared volume ${\cal V}^2$,
given by (minus) the squared of the triple grasping in \Ref{exV},
to obtain a real expectation value and avoid the extra tangent factor.
However, graspings in general do not commute, thus the definition of their products
requires an ordering prescription. For the objective of
studying the semiclassical limit, the appropriate ordering seems to take some sort of
``temporal ordering'': we act with the two triple graspings one after the other, without
allowing them to self--intersect.
With this ordering, it is easy to compute the value of ${\cal V}^2$ starting from the results
listed in Table \ref{table}. In particular, acting twice in a row with the grasping type 2
and proceeding as above, we obtain the following leading order,
\eqa
&&
\left(\f2{l_1} A_{123} A_{156}\right)^2
\left[\left\{\begin{array}{ccc} j_1-2 & j_2 & j_3 \\ j_4 & j_5 & j_6 \end{array} \right\}
-2 \left\{\begin{array}{ccc} j_1 & j_2 & j_3 \\ j_4 & j_5 & j_6 \end{array} \right\}
+ \left\{\begin{array}{ccc} j_1+2 & j_2 & j_3 \\ j_4 & j_5 & j_6 \end{array} \right\}\right]
= \no\no &\simeq&
\f{\left(\f2{l_1} A_{123} A_{156}\right)^2}{\sqrt{12\pi\, V(j_s)}}
\left[\cos\Big(S_{\rm R}(j_1-2)+\f\pi4\Big)-2\cos\Big(S_{\rm R}(j_s)+\f\pi4\Big)
+\cos\Big(S_{\rm R}(j_1+2)+\f\pi4\Big) \right]
= \no\no &\simeq& -4 \f{\left(\f2{l_1} A_{123} A_{156}\right)^2}{\sqrt{12\pi\, V(j_s)}}
\, \sin^2\theta_1 \, \cos\Big(S_{\rm R}(j_s)+\f\pi4\Big)
\equiv -(3!)^2\f{V^2(j_s)}{\sqrt{12\pi\, V(j_s)}}\,\cos\Big(S_{\rm R}(j_s)+\f\pi4\Big),
\neqa
from which we conclude that
\equ\label{limitV2}
\mean{{\cal V}^2(\tilde\tau)} \sim \f1Z \sum_{j\gg1} \prod_s d_{j_s} \prod_{\tau} \{6j\}
\, V_{\tilde\tau}^2(j_s).
\nequ

We obtain the correct semiclassical limit for the value of the squared volume.
Notice that this result can be generalised to arbitrary powers of the triple grasping:
this ordering prescription allows to immediately identify the (semiclassical limit of the)
$n$-th power of the triple
grasping (which is giving raise to $\pm n$ shifts in the \6 symbol) with the $n$-th power
of the volume (times the factor $i$ tangent if $n$ is odd).

In conclusion, the asymptotics of (powers of) the triple grasping
are dominated by (powers of) the classical formula \Ref{V} for the
volume of a tetrahedron. It is interesting to notice that the
polynomial part of this formula arises from the coefficients of the
grasping, see \Ref{lim1}. The non--polynomial part, namely the sine
of the dihedral angle, arises on the other hand from the expansion
of the Regge action entering the modified \6s with $\pm 1$. Thus the
fact that the triple grasping produces a superposition of \6s is
crucial. Furthermore, let us remark again that the fact that the
leading asymptotics come from the terms with $\pm 1$ is somewhat
reminiscent of the fact that in the spin network basis the canonical
3d volume operator in loop quantum gravity has values only on the
superdiagonal and the subdiagonal.

\section{Quantum corrections}\label{SectionCorr}
Having discussed the meaning of the factor $i$ tangent in \Ref{limitV}, we
now look at the next to leading order corrections to the quantum volume.
Let us first consider the quantum corrections to the grasping of type 2. These are the only ones
entering the definition \Ref{volume} of the volume.
There are three types of corrections:
\begin{itemize}
\item{Corrections from $O(l^2)$ terms in the coefficients $c_i(l_s)$. These can be read
from the Appendix, respectively from \Ref{c-exp}, \Ref{c0} and \Ref{c+exp}.
\equ\nonumber
 \f{c_-^{(1)}(j_s)}{\sqrt{12\pi\, V(j_s)}} \, \cos\Big(S_{\rm R}(j_1-1)+\f\pi4\Big)
+ \f{ c_0(j_s)}{\sqrt{12\pi\, V(j_s)}} \, \cos\Big(S_{\rm
R}(j_1)+\f\pi4\Big) +\f{c_+^{(1)}(j_s)}{\sqrt{12\pi\,
V(j_s)}} \, \cos\Big(S_{\rm R}(j_1+1)+\f\pi4\Big). \nequ
Using the fact that $c^{(1)}:=c_-^{(1)} \equiv c_+^{(1)}$ (see the Appendix) and expanding the Regge action
as above, this correction can be simply written as
\equ\label{corr1}
\f i6\f{2\, c^{(1)}(j_s) \, \cos\theta_1+c_0(j_s)}{\sqrt{12\pi\, V(j_s)}} \, \cos\Big(S_{\rm R}(j_s)+\f\pi4\Big).
\nequ
Recall that $c^{(1)}$ and $c_0$ depend on the configuration chosen (\emph{i.e.} the triplet grasped -- here
126). The correction to the volume is obtained summing over all configurations (four different ones using \Ref{volume},
sixteen using \Ref{volume1}), which restores  a symmetric expression.
}

\item{Corrections from higher orders in the expansion \Ref{Rexp} of the Regge action. These
can be obtained as follows. First of all, we keep up to the second order term in the expansion of the Regge action,
\equ\nonumber
S_{\rm R}(j_1\pm 1)\simeq S_{\rm R}(j_s)\pm \theta_1+G_{11}.
\nequ
The exact form of the second derivative $G_{11}:=\f12\f{\p^2 S_{\rm R} }{\p j_1^2}$ can be
computed from elementary geometry (for examples, see \cite{3dcorr}). From dimensional analysis,
it is clear that $G_{11}\sim 1/j$, thus we expect this term to contribute to the corrections.
The contribution can be easily computed. We have
$\sin(\theta_1+G_{11})\simeq \sin\theta_1 +G_{11}\cos\theta_1$
instead of \Ref{trigo}, thus the extra piece gives a correction to \Ref{vol2} of
\equ\label{corr2}
i\, \cot\theta_1 \, G_{11}\, \f{V(j_s)}{\sqrt{12\pi\, V(j_s)}} \, \sin\Big( S_{\rm R}(j_s)+\f\pi4 \Big).
\nequ
Here again, one has to sum over all the relevant configurations to obtain the correction to the volume.
}
\item{Corrections from higher orders in the expansion \Ref{asymp} of the \6 symbol.
Unfortunately, the next term in \Ref{asymp} is not known in the literature, thus we cannot
pursue this analysis to the end. However, numerical investigations performed in \cite{3dcorr}
hint for a next term of the type $\f{N}{V(j_s)^{5/6}}\, \cos(S_{\rm R}(j_s)+\phi)$, where
$N$ and $\phi$ are numerical constants. This ansatz is just one power of $j$ below the first term,
thus it contributes to the next to leading order,
\equ\label{corr3}
\f i6 \, \f{\sqrt{12\pi} \,N}{V(j_s)^{5/3}}\, \f{V(j_s)}{\sqrt{12\pi\, V(j_s)}} \,
\Big[ \sin(\phi-\f\pi4) \, \cos\Big(S_{\rm R}(j_s)+\f\pi4\Big)+
\cos(\phi-\f\pi4) \, \sin\Big(S_{\rm R}(j_s)+\f\pi4\Big) \Big].
\nequ
The relative sign of this correction depends on the sign of the numerical constants $N$ and $\phi$.
The fit obtained in \cite{3dcorr} for the equilateral case when all $j_s$ are the same, gave
$N=-0.36$, $\phi=0.68$.}
\end{itemize}
Summing \Ref{corr1}, \Ref{corr2} and \Ref{corr3} among themselves and over the four configurations
of grasping type 2 (corresponding to grasping the four different vertices), we obtain the
overall  next to leading order correction to the quantum volume.

The analysis of the corrections of the grasping of type 3 leads exactly to the same results
above. However the two definitions \Ref{volume} and \Ref{volume1} do differ at the level of quantum corrections,
as the latter also receives contributions from the degenerate graspings of type 4, 5 and 6.

Consider first the graspings of type 5. Its leading
order can be immediately read from type 2 - provided we take into
account the extra denominators reducing by one power of $j$ the
coefficients of the graspings. We can thus write \equ\label{diff1}
\f i6 \,\f{c_0(j_s)-4{A_{123} A_{156}}
\cos\theta_1/{j_1^2}}{\sqrt{12\pi\, V(j_s)}} \cos\Big(S_{\rm
R}(j_s)+\f\pi4\Big). \nequ

Finally, 4 and 6 can be recognised, using the results of Section \ref{Sectiondouble},
to be the values of the various scalar products between segment vectors, $\ell_s\cdot \ell_{s'}$.
Including all the permutations, we simply have the overall correction
\equ\label{diff}
-\f i6\,\f{\sum_{s,s'} \ell_s\cdot \ell_{s'}}{\sqrt{12\pi\, V(j_s)}}\, \cos\Big(S_{\rm R}(j_s)+\f\pi4\Big).
\nequ

These extra corrections \Ref{diff1} and \Ref{diff} allow us to
distinguish and differentiate between the two definitions of the
quantum volume.

\section{Conclusions}\label{SectionConcl}
By relating the classical observable to a grasping operator and
using the recoupling theory of $\SU(2)$, we explicitly computed the
value of the quantum volume of a labelled tetrahedron in the PR
model. The various terms contributing to the quantum value are
reported in Table \ref{table}, and the details of the evaluation in
the Appendix.

Then, we studied the semiclassical limit by considering the large
spin expansion. The leading order, given in \Ref{limitV}, is indeed
dominated by the classical formula, but an additional factor is
present. This factor, $i$ times the tangent of the Regge action, can
be understood as a consequence of the fact that the PR model sums
over both orientations of spacetime, and was indeed anticipated in
\cite{FK}. Consistently, the value of the squared volume should not
have this problem, and we confirmed this expectation: the value of
the squared volume, given in \Ref{limitV2}, is dominated precisely
by the classical formula. In doing so, we introduced an ordering
prescription for the products of graspings. Using this prescription,
it is easy to see that the $n$th power of the triple grasping has
the asymptotics dominated by the $n$th power of the classical volume
(times a factor $i$ tangent if $n$ is odd). The existence of such an
ordering prescription could turn out to be important in studying the
large spin expansion of spinfoam models of gravity coupled to matter
field.

It is interesting to note that a key factor
of the classical formula, namely the sine of the dihedral angle, does not come directly from the coefficients
of the grasping, but from the fact that the triple grasping changes the initial \6 into a superposition
of \6s. This fact mimics the structure of the canonical volume operator.

We considered two different definitions of the quantum volume, related to 
classically equivalent different discretisations, and the results summarised above
hold for both definitions.
However, we also showed how one can compute the next to leading order quantum
correction, and this is where the two versions of the quantum volume
differ. We expect this to be a generic feature of constructing quantum observables
starting from discrete classical quantities: different quantum observables 
corresponding to classically equivalent quantities, even if
they have the same semiclassical leading order, could still differ at the next to leading order.
The latter can be then used to distinguish them.

In particular, the next to leading order also depends on the second term
in the expansion of the \6 symbol, a term which is not analytically
known at the moment. At this stage, we are thus not yet able to say
what is the relative sign of the correction, and so we leave the
issue open. However, let us stress that knowing the next term of the
\6 is also useful to study how quantum gravity can give rise to a
different perturbative expansion for the graviton propagator, as
suggested in \cite{3dcorr}, so computing this term analytically
could be useful for several reasons.

The results reported here show that the large spin limit reproduces semiclassical geometry
in 3d quantum gravity. It is important to extend a similar analysis to the physically interesting
4d case. However, the situation is more subtle in 4d. In fact, the classical geometrical quantities
do not commute in 4d spinfoam models of quantum gravity. Therefore an eigenstate will not in general
have the semiclassical limit, just as well as one can not study the semiclassical
limit of the harmonic oscillator on an eigenstate of the energy.
It is necessary to first construct appropriate semiclassical states, or coherent states.
A class of semiclassical states for the tetrahedron of Loop Quantum Gravity were
proposed in \cite{semi}. It would be interesting to apply the logic described here
to study the value of the quantum volume of these states.

\section*{Acknowledgments}

We thank Carlo Rovelli for useful discussions on the construction of the volume grasping,
and Laurent Freidel for suggestions on the use of the recoupling theory.

Research at the Perimeter Institute is supported in part by the
Government of Canada through NSERC and by the Province of Ontario
through MEDT.

\appendix

\section{Evaluation of $\SU(2)$ spin networks}
A spin network $s$ is a triple $\{\gamma, j_l, i_n\}$ consisting of
a graph $\gamma$; a set of irreps $j_l$ aassociated to the links; a
set of intertwiners $i_n$ associated to the nodes. A spin network
state $\psi_s(g_l)$ is constructed assigning a group element $g_l$
to each link in the corresponding irrep $j_l$, namely assigning a
representation matrix $D^{(j_l)}(g_l)$, and then contracting the
indices of all the matrices with the intertwiners on the nodes:
\equ\label{spinnet} \psi_s(g_l) = \prod_l D^{(j_l)}(g_l) \ \prod_n
i_n. \nequ By construction, this is a gauge--invariant quantity,
belonging to the space Inv$\Big[\bigotimes_l {\cal H}_{j_l}\Big]$.
Upon the interpretation of the group elements as parallel transports
\Ref{defg}, \Ref{spinnet} becomes a cylindrical function of the
connection \cite{AshtekarMeasure, Baez:1994hx}. The spin networks
span the kinematical Hilbert space of LQG, which is given by \equ
L^2[{\cal A}/{\cal G}]\cong \bigoplus_j \bigotimes_{\rm v} {\rm Inv}
\left[ \bigotimes_l {\cal H}_{j_l}\right]. \nequ

Evaluating a spin network refers to taking all group elements to the
identity, $g_l\mapsto\mathbbm 1$ $\forall l$, and defines a map
$L^2[{\cal A}/{\cal G}]\mapsto \mathbb C$.

\bigskip

To fix the normalisation of the spin networks, consider the $\theta$ graph
\parbox{1.2cm}{\begin{picture}(0,0)(-20,0)\put(0,0){\circle{20}}\Line(-10,0)(10,0)
\put(-15,8){\small$j_1$}\put(0,2){\small$j_2$}\put(-15,-12){\small$j_3$}\end{picture}}.
Using the normalised Wigner $3m$-coefficients as intertwiners, \Ref{spinnet} reads
\equ\label{psitheta}
\psi(g_1, g_2, g_3) = D^{(j_1)}_{m_1n_1}(g_1) D^{(j_2)}_{m_2n_2}(g_2)
D^{(j_3)}_{m_3n_3}(g_3) \left( \begin{array}{ccc} j_1 & j_2 & j_3 \\
m_1 & m_2 & m_3 \end{array} \right)\left( \begin{array}{ccc} j_1 & j_2 & j_3 \\
n_1 & n_2 & n_3 \end{array} \right),
\nequ
and its evaluation gives
\equ
\psi({\mathbbm 1}, {\mathbbm 1}, {\mathbbm 1}) =  \left( \begin{array}{ccc} j_1 & j_2 & j_3 \\
m_1 & m_2 & m_3 \end{array} \right)\left( \begin{array}{ccc} j_1 & j_2 & j_3 \\
m_1 & m_2 & m_3 \end{array} \right)
\equiv \left\{\begin{array}{ll}
1  & {\rm if}  \; |j_1-j_2|\leq j_3 \leq j_1+j_2,     \\  \\ 0 & {\rm otherwise}.
\end{array}\right.
\nequ The inequality that has to be satisfied is the usual
Clebsch--Gordan condition. Given an arbitrary spin network, its
evaluation is identically zero unless the Clebsch--Gordan condition
hold at all nodes. \vspace{0.7cm}

\subsection{The \6 symbol.}
On the tetrahedral graph
\parbox{1.6cm}{\begin{picture}(0,0) (10,10)
\tetnet\end{picture}},
\vspace{0.5cm}
the evaluation of the spin network gives the \6 symbol defined in \Ref{combin6j},
$$
\left\{\begin{array}{ccc} j_1 & j_2 & j_3 \\
j_4 & j_5 & j_6 \end{array} \right\} :=
\left( \begin{array}{ccc} j_1 & j_2 & j_3 \\
m_1 & m_2 & m_3 \end{array} \right)
\left( \begin{array}{ccc} j_1 & j_5 & j_6 \\
m_1 & m_5 & m_6 \end{array} \right)
\left( \begin{array}{ccc} j_4 & j_5 & j_3 \\
m_4 & m_5 & m_3 \end{array} \right)
\left( \begin{array}{ccc} j_4 & j_2 & j_6 \\
m_4 & m_2 & m_6 \end{array} \right).
$$
This object has the symmetries associated with the geometrical symmetries of the tetrahedron pictured above and,
more fundamentally, it satisfies the Biedenharn--Elliott identity,
\equ\label{BH}
\left\{\begin{array}{ccc} j_1 & k_2 & j_3 \\ k_1 & j_2 & k_3  \\ \end{array}\right\}
\left\{\begin{array}{ccc} l_1 & k_2 & l_3 \\ k_1 & l_2 & k_3  \\ \end{array}\right\}
= \sum_j d_j
\left\{\begin{array}{ccc} j_1 & j_2 & k_3 \\ l_2 & l_1 & j  \\ \end{array}\right\}
\left\{\begin{array}{ccc} j_2 & j_3 & k_1 \\ l_3 & l_2 & j  \\ \end{array}\right\}
\left\{\begin{array}{ccc} j_3 & j_1 & k_2 \\ l_1 & l_3 & j  \\ \end{array}\right\}.
\nequ

An analytic expression for the \6 is provided by the Racah formula
\cite{Yutsis}, \equ\label{racah} \left\{\begin{array}{ccc} j_1 & j_2
& j_3 \\ j_4 & j_5 & j_6  \\ \end{array}\right\} =
(-1)^{j_1+j_2+j_3+j_4}\sqrt{\Delta(j_1,j_2,j_3)\Delta(j_1,j_5,j_6)\Delta(j_4,j_5,j_3)\Delta(j_4,j_2,j_6)}
\, \sum_k \f{(-1)^k (k+1)!}{f(k)}, \nequ where \equ
\Delta(j_1,j_2,j_3) := \f{ (j_1+j_2-j_3)! (j_1-j_2+j_3)!
(-j_1+j_2+j_3)!}{(j_1+j_2+j_3+1)!} \nequ and \eqa
f(k)&:=&(k-j_1-j_2-j_3)!(k-j_1-j_5-j_6)!(k-j_4-j_2-j_6)! \times \no
&&\times(k-j_4-j_5-j_3)!(j_1+j_2+j_4+j_5-k)! \times \no
&&\times(j_2+j_3+j_5+j_6-k)!(j_1+j_3+j_4+j_6-k)! \neqa $k$ is an
integer, and the sum in \Ref{racah} is over all admissable $k$. From
the graphical representation of the \6 symbol, it is clear that the
Clebsch--Gordan conditions must hold on all four nodes, so that the
functions $\Delta$ appearing in \Ref{racah} are all well--defined.

Some useful explicit values of \Ref{racah} are the followings:
\eqa\label{un1}
\left\{\begin{array}{ccc} j_1 & j_2 & j_3 \\ 1 & j_3 & j_2  \\ \end{array}\right\}
&=& \f{1}{2}\f{C^2(j_2)+C^2(j_3)-C^2(j_1) }{\sqrt{d_{j_2} \ C^2(j_2)\ d_{j_3} \ C^2(j_3)}},
\\ \no\no
\left\{\begin{array}{ccc} j_1 & j_2 & j_3 \\ 1 & j_3 & j_2+1  \\
\end{array}\right\} &=& \f12
\frac{\sqrt{(1+j_2+j_3 -j_1)(1+j_2-j_3
+j_1)(-j_2+j_3+j_1)(2+j_2+j_3+j_1)}}{\sqrt{(j_2+1) \, d_{j_3} \, d_{j_2} \, d_{j_2+1} \, C^2(j_3)}},
\\ \no\no
\left\{\begin{array}{ccc} j_1 & j_2 & j_3 \\ 1 & j_3 & j_2-1
\\ \end{array}\right\} &=& -\f12\frac{\sqrt{(j_2+j_3 -j_1)(j_2-j_3
+j_1)(1-j_2+j_3+j_1)(1+j_2+j_3+j_1)}}{\sqrt{j_2\, d_{j_3} \, d_{j_2} \, d_{j_2-1} \, C^2(j_3)}},
\neqa
from which we immediately read
\eqa
\left\{\begin{array}{ccc} j & j & 1 \\ 1 & 1 & j
\\ \end{array}\right\} = \frac{{1}}{\sqrt{6 \, d_{j} \, C^2(j)}},
\qquad
\left\{\begin{array}{ccc} j & j & 1 \\ 1 & 1 & j+1
\\ \end{array}\right\} = \frac{j}{\sqrt{6 \, d_{j} \, C^2(j)}},
\qquad
\left\{\begin{array}{ccc} j & j & 1 \\ 1 & 1 & j-1
\\ \end{array}\right\} = -\frac{j+1}{\sqrt{6 \, d_{j}\, C^2(j)}}.
\neqa

As it will be useful in the following, let us recall here the definition of the \{9j\} symbol,
\equ\label{def9j}
\left\{\begin{array}{ccc} j_1 & j_2 & j_3 \\ j_4 & j_5 & j_6  \\ j_7 & j_8 & j_9  \\ \end{array}\right\}=
\sum_j \,d_j\;
\left\{\begin{array}{ccc} j_1 & j_2 & j_3 \\ j_6 & j_9 & j  \\ \end{array}\right\}
\left\{\begin{array}{ccc} j_4 & j_5 & j_6 \\ j_2 & j & j_8  \\ \end{array}\right\}
\left\{\begin{array}{ccc} j_7 & j_8 & j_9 \\ j & j_1 & j_4  \\ \end{array}\right\}.
\nequ
Notice that it is antisymmetric under exchange of rows, thus
\equ\label{9j0}
\left\{\begin{array}{ccc} j_1 & j_2 & j_3 \\ j_1 & j_2 & j_3  \\ j_7 & j_8 & j_9  \\ \end{array}\right\}\equiv 0.
\nequ
The \{9j\} symbol is the evaluation of the spin network on the hexagonal graph \hspace{0.3cm}
\parbox{0.4cm}{\setlength{\unitlength}{1 pt}\begin{picture}(0,0)
                \put(0,0){\circle{20}}
        \put(-10,0){\line(1,0){20}}\put(-7,-7){\line(1,1){14}}\put(-7,7){\line(1,-1){14}}
        \end{picture}}.

\section{Grasping rules and recoupling theory}
The action of an algebra derivative on a group element in a representation $j$ is given by
\equ
\f\d{\d J^I} D^{(j)}(e^{J})\Big|_{J=0} = -iT^{I(j)},
\nequ
where $T^{I(j)}$ is the $I$-th generator, in the irrep $j$.
We picture this action as $(-i)$\parbox{0.9cm}{
\begin{picture}(0,0)\Line(0,0)(20,0)\DashLine(10,0)(10,10){2}\put(20,-5){\small{$j$}}
\end{picture}}.
The dashed line represents the insertion of the algebra generator.
A dashed 3-valent vertex \hspace{0.2cm} \parbox{0.3cm}{\begin{picture}(0,0)\DashLine(0,0)(0,6){2}\DashLine(0,0)(6,-4){2}\DashLine(0,0)(-6,-4){2}
\end{picture}}
represents the algebra structure constants $i\eps_{ijk}$.
The insertion of an algebra generator is equivalent to adding a link in the adjoint
irrep, up to normalisation. We fix the normalisation\footnote{For a different normalisation, see for instance  \cite{DePietriScaling}}
as to match the grasping rules of Bar--Natan \cite{Bar-Natan}, namely
\equ
\parbox{0.4cm}{\begin{picture}(0,0)\DashLine(0,-10)(0,10){2}\put(0,0){\circle{20}}\end{picture}}=
-\hspace{0.4cm}\parbox{0.4cm}{\begin{picture}(0,0)\DashLine(0,0)(0,10){2}\DashLine(0,0)(8,-6){2}\DashLine(0,0)(-8,-6){2}
\put(0,0){\circle{20}}\end{picture}}
= C^2(j) \hspace{0.4cm} \parbox{0.4cm}{\begin{picture}(0,0)\put(0,0){\circle{20}}\end{picture}}.
\nequ
To this end, we choose
\equ
\parbox{0.9cm}{
\begin{picture}(0,0)\Line(0,0)(20,0)\DashLine(10,0)(10,10){2}\put(20,-5){\small{$j$}}
\end{picture}} = \sqrt{d_j \, C^2(j)}\ \parbox{0.9cm}{
\begin{picture}(0,0)\Line(0,0)(20,0)\Line(10,0)(10,10)\put(12,8){\small{$1$}}\put(20,-5){\small{$j$}}
\end{picture}}, \hspace{2cm}
\parbox{0.4cm}{\begin{picture}(0,0)\DashLine(0,0)(0,10){2}\DashLine(0,0)(8,-6){2}\DashLine(0,0)(-8,-6){2}
\end{picture}} = -\sqrt{6} \hspace{0.5cm} \parbox{0.6cm}{\begin{picture}(0,0)\Line(0,0)(0,10)\Line(0,0)(8,-6)\Line(0,0)(-8,-6)
\put(2,8){\small{$1$}}\put(10,-10){\small{$1$}}\put(-4,-10){\small{$1$}}\end{picture}}.
\nequ

With this normalization, the action of the double grasping is the
insertion in the spin network of an additional link in the adjoint
irrep, weighted by the labels of the two links grasped:
\vspace{0.3cm} \equ\label{doppio} -\f{\d}{\d J^I_1}\f{\d}{\d
J^I_{2}} \Big(
\parbox{1.5cm}{\setlength{\unitlength}{1 pt}\begin{picture}(0,0)
                \put(10,0){\circle{10}}\put(30,0){\circle{10}}
        \put(10,-20){\line(0,1){15}}\put(10,5){\line(0,1){15}}\put(30,-20){\line(0,1){15}}\put(30,5){\line(0,1){15}}
        \put(12,-15){\small$j_1$}\put(32,-15){\small$j_2$}\put(6,-2){$\scriptstyle{J_1}$}\put(26,-2){$\scriptstyle{J_2}$}
        \end{picture}}\Big)\Big|_{J=0}
\equiv \Big(
\parbox{1.5cm}{\setlength{\unitlength}{1 pt}\begin{picture}(0,0)
\DashLine(10,0)(30,0){2}
\put(10,-20){\line(0,1){40}}\put(30,-20){\line(0,1){40}}
        \put(12,-15){\small$j_1$}\put(32,-15){\small$j_2$}
        \end{picture}}\Big)=
\sqrt{d_{j_1} \, C^2(j_1) \, d_{j_2} \, C^2(j_2)}
\Big(
\parbox{1.5cm}{\setlength{\unitlength}{1 pt}\begin{picture}(0,0)
\put(10,0){\line(1,0){20}}
\put(10,-20){\line(0,1){40}}\put(30,-20){\line(0,1){40}}
        \put(12,-15){\small$j_1$}\put(32,-15){\small$j_2$}
        \end{picture}}\Big).
\nequ
\vspace{0.3cm}

\noindent
Here and in the following, a line with no labels means a link coloured
with the adjoint $j=1$ irrep.
In the same way, the triple grasping is the insertion of a 3-valent node
in the adjoint irrep, weighted by the labels of the three links grasped:
\vspace{0.3cm}
\eqa\label{triplo}
\eps^{IJK}\f{\d}{\d J^I_{1}}\f{\d}{\d J^J_{2}}\f{\d}{\d J^K_{3}} \Big(
\parbox{2.3cm}{\setlength{\unitlength}{1 pt}\begin{picture}(0,0)
                \put(10,0){\circle{10}}\put(30,0){\circle{10}}\put(50,0){\circle{10}}
        \put(10,-20){\line(0,1){15}}\put(10,5){\line(0,1){15}}\put(30,-20){\line(0,1){15}}\put(30,5){\line(0,1){15}}
        \put(50,-20){\line(0,1){15}}\put(50,5){\line(0,1){15}}
        \put(12,-15){\small$j_1$}\put(32,-15){\small$j_2$}\put(6,-2){$\scriptstyle{J_1}$}\put(26,-2){$\scriptstyle{J_2}$}
        \put(52,-15){\small$j_3$}\put(46,-2){$\scriptstyle{J_3}$}
        \end{picture}}\Big)\Big|_{J=0}
&\equiv& \Big(
\parbox{2.3cm}{\setlength{\unitlength}{1 pt}\begin{picture}(0,0)
        \put(10,-20){\line(0,1){40}}\put(30,-20){\line(0,1){40}}\put(50,-20){\line(0,1){40}}
        \put(12,-15){\small$j_1$}\put(32,-15){\small$j_2$}\put(52,-15){\small$j_3$}
        \DashLine(10,5)(20,5){2}\DashLine(20,5)(30,10){2}\DashLine(20,5)(50,-5){2}
  \end{picture}}\Big)= \no\no\no &=&
  -\sqrt{6 \, d_{j_1} \, C^2(j_1) \, d_{j_2} \, C^2(j_2) \, d_{j_3} \, C^2(j_3)}
  \Big(
\parbox{2.3cm}{\setlength{\unitlength}{1 pt}\begin{picture}(0,0)
        \put(10,-20){\line(0,1){40}}\put(30,-20){\line(0,1){40}}\put(50,-20){\line(0,1){40}}
        \put(12,-15){\small$j_1$}\put(32,-15){\small$j_2$}\put(52,-15){\small$j_3$}
        \Line(10,5)(20,5)\Line(20,5)(30,10)\Line(20,5)(50,-5)
  \end{picture}}\Big).
\neqa
\vspace{0.3cm}

Therefore, evaluating the action of a grasping operator on a given
spin network amounts to evaluating a new spin network, obtained by
adding to the former the graph of the grasp in the adjoint irrep. To
compute the action of the grasping operators, it is then convenient
the use of the recoupling theory.

The key equation to be used is the recoupling theorem:

\equ
\framebox{\parbox{10cm}{
\begin{eqnarray}
\begin{array}{c}\setlength{\unitlength}{1 pt}
\begin{picture}(50,40)
          \put( 0,0){$j_1$}\put( 0,30){$j_2$}
          \put(45,0){$j_4$}\put(45,30){$j_3$}
          \put(10,10){\line(1,1){10}}\put(10,30){\line(1,-1){10}}
          \put(30,20){\line(1,1){10}}\put(30,20){\line(1,-1){10}}
          \put(20,20){\line(1,0){10}}\put(22,25){$i$}
          \put(20,20){\circle*{3}}\put(30,20){\circle*{3}}
\end{picture}\end{array}
    &=& \sum_k  {\rm dim}\, k \;\left\{\begin{array}{ccc}
                      j_1  & j_2 & i \\
                      j_3  & j_4 & k
                \end{array}\right\}
\begin{array}{c}\setlength{\unitlength}{1 pt}
\begin{picture}(40,40)
      \put( 0,0){$j_1$}\put( 0,40){$j_2$}
      \put(35,0){$j_4$}\put(35,40){$j_3$}
      \put(10,10){\line(1,1){10}}\put(10,40){\line(1,-1){10}}
      \put(20,30){\line(1,1){10}}\put(20,20){\line(1,-1){10}}
      \put(20,20){\line(0,1){10}}\put(22,22){$k$}
      \put(20,20){\circle*{3}}\put(20,30){\circle*{3}}
\end{picture}\end{array}
\nonumber\neqa}} \label{eq:recTheorem} \nequ \noindent From the
definition of the $\theta$ spin network and the Clebsch--Gordan
conditions, it follows that

\vspace{0.3cm}
\eqa\label{bubble}
\begin{array}{c}
\parbox{0.8cm}{\begin{picture}(0,0)
\put(0,0){\circle{20}}\put(0,-10){\circle*{3}}\put(0,10){\circle*{3}}
\put(0,10){\line(0,1){20}}\put(0,-10){\line(0,-1){20}}
\put(-8,20){$j_1$}\put(-18,-2){$j_2$}\put(12,-2){$j_3$}\put(-8,-25){$j_4$}
\end{picture}}\end{array}
=\f{1}{{\rm dim}\,j_1}\, \d_{j_1j_4}\quad
\parbox{0.4cm}{\begin{picture}(0,0)
\put(0,-30){\line(0,1){60}}\put(-8,20){$j_1$}\end{picture}}.
\neqa
\vspace{0.3cm}

Using \Ref{eq:recTheorem} repeatedly, it is easy to demonstrate the following 3-vertex contraction:
\begin{equation}
\begin{array}{c}
  \setlength{\unitlength}{1 pt}
  \begin{picture}(55,50)
       \put(25, 5){\line(0,1){10}}   \put(27,1){$j_1$}
       \put(25,15){\circle*{3}}
       \put(25,15){\line(-1, 1){10}} \put(12,14){${j_6}$}
       \put(25,15){\line( 1, 1){10}} \put(32,14){${j_5}$}
       \put(35,25){\circle*{3}}\put(15,25){\circle*{3}}
       \put(15,25){\line(-1, 1){10}} \put( 4,37){$j_2$}
       \put(35,25){\line( 1, 1){10}} \put(42,37){$j_3$}
       \put(15,25){\line(1,0){20}}   \put(25,29){$j_4$}
\end{picture}\end{array}
 = {\left\{\begin{array}{ccc} j_1 &  j_2  & j_3 \\ j_4 & j_5 & j_6  \end{array}\right\}}
\begin{array}{c}\setlength{\unitlength}{1 pt}
\begin{picture}(40,40)
       \put(15,15){\line(-1, 1){10}} \put( 4,27){$j_2$}
       \put(15,15){\line( 1, 1){10}} \put(22,27){$j_3$}
       \put(15, 5){\line(0,1){10}}   \put(17,1){$j_1$}
       \put(15,15){\circle*{3}}
\end{picture}\end{array}
~,
\label{3vertexReduction}
\end{equation}
Notice that if we add an extra node to close the links $j_1$, $j_2$ and $j_3$  on both sides of
\Ref{3vertexReduction}, and using the normalisation of the $\theta$ spin network,
we get back the definition of the \6 symbol as the evaluation of the
tetrahedral spin network.

To evaluate the graspings, the strategy is the following:
we use the definitions \Ref{doppio} and \Ref{triplo} to obtain new graphs; then we use
\Ref{eq:recTheorem} enough times to reduce the new graphs to the original ones. In particular
in the next Sections, we will apply this strategy to the tetrahedral graph, in order to study
the spectra of geometrical observables in the Ponzano--Regge model.

\subsection{The double grasping on the \6}
Here we show how to evaluate the double grasping on the tetrahedral spin network.
These results were used in Section \ref{Sectiondouble}.

Computing the quadratic grasping on a single link is trivial, all we need is the
bubble move \Ref{bubble}:

\eqa\label{double}
-\frac{\d}{\d J_{s_1}^I} \frac{\d}{\d J_{s_1}^I}
\parbox[2cm]{1.2cm}{\begin{picture}(0,0) (10,15) \sorgenti \end{picture}}\;\;\Bigg|_{J=0}
=
\parbox[2cm]{1.8cm}{\begin{picture}(0,0) (10,9)\tetnet\SetWidth{3}\DashCArc(352,80)(40,-100,100){8}\end{picture}}
=
d_{j_1} \, C^2(j_1)
\Bigg(\parbox[2cm]{1.8cm}{\begin{picture}(0,0) (10,9)\tetnet\SetWidth{3}\CArc(352,80)(40,-100,100)\end{picture}}\Bigg)
= C^2(j_1) \, \{6j\}.
\neqa
This proves \Ref{GraspingCas} used to compute the spectrum of lengths.

Computing the grasping on two different segments sharing a point is also very simple.
We have

\eqa\label{double1}
-\frac{\d}{\d J_{s_1}^I} \frac{\d}{\d J_{s_2}^I}
\parbox[2cm]{1.2cm}{
\begin{picture}(0,0) (10,15) \sorgenti \end{picture}}\;\;\Bigg|_{J=0}
= \parbox[2cm]{1.6cm}{\begin{picture}(0,0) (10,9)
\tetnet \SetWidth{3}\DashCArc(210.86,139.15)(38.63,-38.7,124.46){8}
\end{picture}}
= \sqrt{d_{j_1} \, C^2(j_1) \, d_{j_2} \, C^2(j_2) }
\Bigg(\parbox[2cm]{1.8cm}{\begin{picture}(0,0) (10,9)\tetnet \SetWidth{3}\CArc(210.86,139.15)(38.63,-38.7,124.46)
\end{picture}}\Bigg).
\neqa
The graph obtained with the grasping can be straightforwardly reduced to the initial
tetrahedral graph, simply using \Ref{3vertexReduction} once:
\eqa
\parbox[2cm]{1.6cm}{\begin{picture}(0,0) (10,9)
\tetnet \SetWidth{3}\CArc(210.86,139.15)(38.63,-38.7,124.46)
\end{picture}}
= \left\{\begin{array}{ccc} j_3 & j_1 & j_2 \\ 1 & j_2 & j_1  \\ \end{array}\right\}
\Bigg(\ \parbox[2cm]{1.6cm}{\begin{picture}(0,0) (15,9)
\tetnet\end{picture}}\Bigg),
\neqa

\noindent Therefore, using the explicit expression \Ref{un1}, we obtain \Ref{Grasping2},
\eqa
-\frac{\d}{\d J_{s_1}^I} \frac{\d}{\d J_{s_2}^I}
\parbox[2cm]{1.6cm}{\fcolorbox{white}{white}{
\begin{picture}(0,0) (10,15) \sorgenti \end{picture}}}\;\;\Bigg|_{J=0} =
\f12 [C^2(j_1) + C^2(j_2) -C^2(j_3)] \, \{6j\},
\neqa
which we used to compute the spectrum of angles.

\subsection{The triple grasping on the \6}\label{AppTriplo}
Here we show how to evaluate the triple grasping on the \6 symbol, a
result used in Section \ref{Sectiontriple}. As shown in Table
\ref{table}, there are several types of graspings that one has to
consider. We proceed in the order given there.

\bigskip

\emph{Grasping 1.}
\eqa
&& \left(\epsilon_{IJK} \frac{\d}{\d J_{s_1}^I} \frac{\d}{\d J_{s_2}^J} \frac{\d}{\d J_{s_3}^K} \right)
\parbox[2cm]{1.2cm}{\begin{picture}(0,0) (10,15) \sorgenti \end{picture}
}\;\;\Bigg|_{J=0} =
\parbox{1.6cm}{\begin{picture}(0,0) (10,9)\tetnet
\SetColor{Blue}\DashLine(190,110)(345,110){8}\DashLine(190,110)(160,135){8}\DashLine(190,110)(220,135){8}\end{picture}} =
\no\no\no\no && \hspace{-1cm}
= - \sqrt{6 \, d_{j_1} \, C^2(j_1) \, d_{j_2} \, C^2(j_2) \, d_{j_3} \, C^2(j_3)} \sum_k d_k \;
\left\{\begin{array}{ccc} j_2 & j_1 & j_3 \\ 1 & j_3 & k  \\ \end{array}\right\}
\left\{\begin{array}{ccc} j_3 & j_1 & j_2 \\ 1 & j_2 & k  \\ \end{array}\right\}
\left\{\begin{array}{ccc} j_1 & j_1 & 1 \\ 1 & 1 & k  \\ \end{array}\right\}
\left\{\begin{array}{ccc} j_1 & j_2 & j_3 \\ j_4 & j_5 & j_6  \\ \end{array}\right\}.
\neqa
If we use the symmetries of the \6 symbol, we see that the coefficient obtained in front of the original \6 matches the definition of the $\{9j\}$ symbol introduced in \Ref{def9j},
\equ
\sum_k d_k \;
\left\{\begin{array}{ccc} j_1 & j_2 & j_3 \\ j_3 & 1 & k  \\ \end{array}\right\}
\left\{\begin{array}{ccc} j_1 & j_2 & j_3 \\ j_2 & k & 1  \\ \end{array}\right\}
\left\{\begin{array}{ccc} 1 & 1 & 1 \\ k & j_1 & j_1  \\ \end{array}\right\}
= \left\{\begin{array}{ccc} j_1 & j_2 & j_3 \\ j_1 & j_2 & j_3  \\ 1 & 1 & 1  \\ \end{array}\right\}
\equiv 0.
\nequ
By symmetry, this result applies to all coplanar triples, and we recover the classical
result: there is no contribution to the volume from coplanar segments.

\bigskip

\emph{Grasping 2.}
\eqa\label{Gras2}
&& \left(\epsilon_{IJK} \frac{\d}{\d J_{s_1}^I} \frac{\d}{\d J_{s_2}^J} \frac{\d}{\d J_{s_6}^K} \right)
\parbox[2cm]{1.2cm}{\begin{picture}(0,0) (10,15) \sorgenti \end{picture}
}\;\;\Bigg|_{J=0} =
\parbox{1.6cm}{\begin{picture}(0,0) (10,9)\tetnet
\SetColor{Red}\DashLine(290,80)(346,80){8}\DashLine(290,80)(245,110){8}\DashLine(290,80)(245,50){8}\end{picture}} =
\no\no\no\no && \hspace{-1cm}
= - \sqrt{6 \, d_{j_1} \, C^2(j_1) \, d_{j_2} \, C^2(j_2) \, d_{j_6} \, C^2(j_6)} \sum_k d_k \;
\left\{\begin{array}{ccc} j_3 & j_1 & j_2 \\ 1 & j_2 & k  \\ \end{array}\right\}
\left\{\begin{array}{ccc} j_5 & j_1 & j_6 \\ 1 & j_6 & k  \\ \end{array}\right\}
\left\{\begin{array}{ccc} j_1 & j_1 & 1 \\ 1 & 1 & k  \\ \end{array}\right\}
\left\{\begin{array}{ccc} k & j_2 & j_3 \\ j_4 & j_5 & j_6  \\ \end{array}\right\}.
\neqa
The calculation yields substantial differences from the previous one. The main one is the
new spin $k$ replacing the original $j_1$ in the \6 (by symmetry, we could have replaced $j_2$
or $j_6$, without changing the final result). From the third \6 in the coefficient above, we read that
the sum ranges over $k=j_1-1, j_1, j_1+1$. We can thus write the result above as
\equ
c_-(j_s) \left\{\begin{array}{ccc} j_1-1 & j_2 & j_3 \\
j_4 & j_5 & j_6 \end{array} \right\} + c_0(j_s) \left\{\begin{array}{ccc} j_1 & j_2 & j_3 \\
j_4 & j_5 & j_6 \end{array} \right\} + c_+(j_s) \left\{\begin{array}{ccc} j_1+1 & j_2 & j_3 \\
j_4 & j_5 & j_6 \end{array} \right\},
\nequ
where, using the formulae reported in the Appendix A,
\eqa
c_-(j_s) &=& \frac{(j_1 + 1)}{4 j_1(2j_1 + 1)}
\sqrt{(j_1+j_2-j_3)(j_1-j_2+j_3)(1-j_1+j_2+j_3)(1+j_1+j_2+j_3)}\times \no\no
&& \hspace{1cm} \sqrt{(j_1+j_5-j_6)(j_1-j_5+j_6)(1-j_1+j_5+j_6)(1+j_1+j_5+j_6)},
\\\no\label{c0}
c_0(j_s) &=& - \frac{[C^2(j_1) +C^2(j_2) - C^2(j_3)][C^2(j_1) +C^2(j_6) -C^2(j_5)]}{4\, C^2(j_1)},
\\\no
c_+(j_s) &=& - \frac{j_1}{4(j_1 +1)(2j_1 +1)}
\sqrt{(1+j_1+j_2-j_3)(1+j_1-j_2+j_3)(-j_1+j_2+j_3)(2+j_1+j_2+j_3)
}\times \no\no &&
\hspace{1cm}\sqrt{(1+j_1+j_5-j_6)(1+j_1-j_5+j_6)(-j_1+j_5+j_6)(2+j_1+j_5+j_6)}.
\neqa Let us now discuss the large spin expansion of this result.
First of all, recall that the spins are related to the (here adimensional) lengths
through $l_s = j_s + \frac{1}{2}$.
Second, let us recall Heron's formula \cite{heron} for the area of a
triangle with side lengths $a$, $b$, $c$: \equ A_{abc}=\frac{1}{4}
\sqrt{(a+b+c)(-a+b+c)(a-b+c)(a+b-c)}. \nequ As a useful shorthand,
we will define the quantity \equ V_1 = \frac{2}{l_1}
A_{l_1l_2l_3}A_{l_1l_5l_6}, \nequ and generalize to arbitrary $V_i$
by the symmetry of the tetrahedron. Notice that from \Ref{V} we
immediately have $V = \f13V_i \sin\theta_i$.

The coefficients of the functions in the lengths $c_\pm(l_s)$ are
polynomials of order three. To second order, we have
\eqa\label{c-exp} c_-(l_s) &\simeq& V_1 \Bigg[\left(1
+\frac{1}{l_1}\right)\left(1
-\frac{1}{4(l_1+l_2-l_3)}-\frac{1}{4(l_1-l_2+l_3)}
+\frac{1}{4(-l_1+l_2+l_3)}-\frac{1}{4(l_1+l_2+l_3)} \right) \times
\no && \hspace{1cm} \left(1
-\frac{1}{4(l_1+l_5-l_6)}-\frac{1}{4(l_1-l_5+l_6)}
+\frac{1}{4(-l_1+l_5+l_6)}-\frac{1}{4(l_1+l_5+l_6)} \right)
+o\left(\frac{1}{l^2} \right)\Bigg],
\\\no\label{c+exp}
c_+(l_s) &\simeq& - V_1\Bigg[\left(1 - \frac{1}{l_1}  \right)
\left(1
+\frac{1}{4(l_1+l_2-l_3)}+\frac{1}{4(l_1-l_2+l_3)}-\frac{1}{4(-l_1+l_2+l_3)}+\frac{1}{4(l_1+l_2+l_3)}\right)
\times \no && \hspace{1cm} \left(1+\frac{1}{4(l_1+l_5-l_6)}+\frac{1}{4(l_1-l_5+l_6)}-
\frac{1}{4(-l_1+l_5+l_6)}+\frac{1}{4(l_1+l_5+l_6)}\right)
+o\left(\frac{1}{l^2} \right) \Bigg] \neqa
Let us define the coefficients of the expansion
as in $c_{\pm}(j_s)=c_{\pm}^{(0)}(j_s) + c_{\pm}^{(1)}(j_s)+\ldots$
From the equations above, we see that $c_{\pm}^{(0)}(j_s)=\mp V_1$, and $c_{-}^{(1)}(j_s)\equiv c_{+}^{(1)}(j_s)$.

All terms in $c_0(j_s)$, on the other hand, are of order $j^2$.
In particular, notice that
$c_0(j_s) = - (\ell_1\cdot\ell_2) \, (\ell_1\cdot\ell_6) / \ell_1^2$.

The values from the other three vertex graspings can be obtained by symmetry.
The other three contributions can be obtained as above,
considering the other three cases $(j_1, j_3, j_5)$, $(j_2, j_3, j_4)$ and
$(j_4, j_5, j_6)$.

\bigskip

\emph{Grasping 3.}
\eqa\label{Gras3}
&& \left(\epsilon_{IJK} \frac{\d}{\d J_{s_1}^I} \frac{\d}{\d J_{s_2}^J} \frac{\d}{\d J_{s_4}^K} \right)
\parbox[2cm]{1.2cm}{\begin{picture}(0,0) (10,15) \sorgenti \end{picture}
}\;\;\Bigg|_{J=0} =
\parbox{1.6cm}{\begin{picture}(0,0) (10,9)\tetnet
\SetColor{Red}\DashLine(190,110)(345,110){8}\DashLine(190,110)(190,80){8}\DashLine(190,110)(220,135){8}\end{picture}} =
\no\no\no\no && \hspace{-1cm}
= - \sqrt{6 \, d_{j_1} \, C^2(j_1) \, d_{j_2} \, C^2(j_2) \, d_{j_4} \, C^2(j_4)} \sum_k d_k \;
\left\{\begin{array}{ccc} j_3 & j_2 & j_1 \\ 1 & j_1 & k  \\ \end{array}\right\}
\left\{\begin{array}{ccc} j_6 & j_2 & j_4 \\ 1 & j_4 & k  \\ \end{array}\right\}
\left\{\begin{array}{ccc} j_2 & j_2 & 1 \\ 1 & 1 & k  \\ \end{array}\right\}
\left\{\begin{array}{ccc} j_1 & k & j_3 \\ j_4 & j_5 & j_6  \\ \end{array}\right\}.
\neqa
Under the permutation $123456\mapsto 213645$, \Ref{Gras3} is equivalent to \Ref{Gras2}.
Therefore, this grasping reproduces the same results as the previous
one. Notice that this is in agreement with the classical result,
\equ
V=\f1{3!}\, e_1\w e_2\w e_6 = \f1{3!} \, e_1\w e_2\w e_4.
\nequ

\bigskip

\emph{Grasping 4.}
\eqa\label{Gras4}
&& \left(\epsilon_{IJK} \frac{\d}{\d J_{s_1}^I} \frac{\d}{\d J_{s_1}^J} \frac{\d}{\d J_{s_2}^K} \right)
\parbox[2cm]{1.2cm}{\begin{picture}(0,0) (10,15) \sorgenti \end{picture}
}\;\;\Bigg|_{J=0} =
\parbox{1.6cm}{\begin{picture}(0,0) (10,9)\tetnet
\SetColor{Blue}\DashLine(250,100)(300,100){8}\DashLine(300,100)(345,70){8}\DashLine(300,100)(345,130){8}\end{picture}}
= \no\no\no\no && \hspace{-1cm}
= - \sqrt{6  \, d_{j_2} \, C^2(j_2)}\, d_{j_1} \, C^2(j_1)
\left\{\begin{array}{ccc} j_3 & j_1 & j_2 \\ 1 & j_2 & j_1  \\ \end{array}\right\}
\left\{\begin{array}{ccc} j_1 & j_1 & 1 \\ 1 & 1 & j_1  \\ \end{array}\right\}
\left\{\begin{array}{ccc} j_1 & j_2 & j_3 \\ j_4 & j_5 & j_6  \\ \end{array}\right\}
= \no &=& -\f12\left[C^2(j_1)+C^2(j_2)-C^2(j_3) \right]
\left\{\begin{array}{ccc} j_1 & j_2 & j_3 \\ j_4 & j_5 & j_6  \\ \end{array}\right\}.
\neqa

\bigskip

\emph{Grasping 5.}
\eqa\label{Gras5}
&& \left(\epsilon_{IJK} \frac{\d}{\d J_{s_1}^I} \frac{\d}{\d J_{s_1}^J} \frac{\d}{\d J_{s_4}^K} \right)
\parbox[2cm]{1.2cm}{\begin{picture}(0,0) (10,15) \sorgenti \end{picture}
}\;\;\Bigg|_{J=0} =
\parbox{1.6cm}{\begin{picture}(0,0) (10,9)\tetnet
\SetColor{Blue}\DashLine(190,80)(280,100){8}\DashLine(280,100)(345,70){8}\DashLine(280,100)(345,130){8}\end{picture}}
= \no\no\no\no && \hspace{-1cm}
- \sqrt{6 \, d_{j_1} \, C^2(j_1) \, d_{j_1} \, C^2(j_1) \, d_{j_4} \, C^2(j_4)} \sum_k d_k \;
\left\{\begin{array}{ccc} j_2 & j_6 & j_4 \\ 1 & j_4 & k  \\ \end{array}\right\}
\left\{\begin{array}{ccc} j_5 & j_6 & j_1 \\ 1 & j_1 & k  \\ \end{array}\right\}
\left\{\begin{array}{ccc} j_1 & j_1 & 1 \\ 1 & 1 & j_1  \\ \end{array}\right\}
\left\{\begin{array}{ccc} j_1 & j_2 & j_3 \\ j_4 & j_5 & k  \\ \end{array}\right\}
= \no &=& \f{c_-(j_s)}{j_1+1} \left\{\begin{array}{ccc} j_1-1 & j_2 & j_3 \\ j_4 & j_5 & j_6 \end{array} \right\} + c_0(j_s) \left\{\begin{array}{ccc} j_1 & j_2 & j_3 \\ j_4 & j_5 & j_6 \end{array} \right\}
+ \f{c_+(j_s)}{j_1} \left\{\begin{array}{ccc} j_1+1 & j_2 & j_3 \\ j_4 & j_5 & j_6 \end{array} \right\}.
\neqa

\bigskip

\emph{Grasping 6.}
\eqa\label{Gras6}
&& \left(\epsilon_{IJK} \frac{\d}{\d J_{s_1}^I} \frac{\d}{\d J_{s_1}^J} \frac{\d}{\d J_{s_1}^K} \right)
\parbox[2cm]{1.2cm}{\begin{picture}(0,0) (10,15) \sorgenti \end{picture}
}\;\;\Bigg|_{J=0} =
\parbox{1.6cm}{\begin{picture}(0,0) (10,9)\tetnet
\SetColor{Blue}\DashLine(345,100)(300,100){8}\DashLine(300,100)(345,70){8}\DashLine(300,100)(345,130){8}\end{picture}}
= \no\no\no\no && \hspace{-1cm}
= - \sqrt{6} \, \left[d_{j_1} \, C^2(j_1)\right]^{\f32} \f1{d_{j_1}}
\left\{\begin{array}{ccc} j_1 & j_1 & 1 \\ 1 & 1 & j_1  \\ \end{array}\right\}
\left\{\begin{array}{ccc} j_1 & j_2 & j_3 \\ j_4 & j_5 & j_6  \\ \end{array}\right\}
= -C^2(j_1)\, \left\{\begin{array}{ccc} j_1 & j_2 & j_3 \\ j_4 & j_5 & j_6  \\ \end{array}\right\}.
\neqa


\begin{thebibliography}{10}

\bibitem{carlo}
C.~Rovelli.  \newblock {\em Quantum Gravity}.  \newblock (Cambridge
University Press, Cambridge 2004.)


\bibitem{Ponzano}
G.~Ponzano, T.~Regge.  ``Semiclassical limit of
Racah coefficients", in {\em Spectroscopy and group theoretical methods
in Physics}, F.~Bloch ed. (North-Holland, Amsterdam, 1968).

\bibitem{FreidelSFM}
L.~Freidel and K.~Krasnov,
``Spin foam models and the classical action principle,''
Adv.\ Theor.\ Math.\ Phys.\  {\bf 2} (1999) 1183
[arXiv:hep-th/9807092].

\bibitem{FK}
  L.~Freidel and K.~Krasnov,
  ``Discrete space-time volume for 3-dimensional BF theory and quantum
  gravity,''
  Class.\ Quant.\ Grav.\  {\bf 16} (1999) 351
  [arXiv:hep-th/9804185].

\bibitem{Fairbairn}
  W.~Fairbairn,
  ``Fermions in three-dimensional spinfoam quantum gravity,''
  arXiv:gr-qc/0609040.

\bibitem{YM}
L. Freidel, C.~Rovelli and S.~Speziale,
``Perturbative expansion for 3d Yang--Mills theory coupled to quantum
gravity in the spinfoam formalism,'' in preparation.

\bibitem{graviton}
  C.~Rovelli,
  ``Graviton propagator from background-independent quantum gravity,''
Phys.\ Rev.\ Lett.\  {\bf 97} (2006) 151301
  [arXiv:gr-qc/0508124].
\\
 E.~Bianchi, L.~Modesto, C.~Rovelli and S.~Speziale,
  ``Graviton propagator in loop quantum gravity,''
    Class.\ Quant.\ Grav.\  {\bf 23} (2006) 6989
  [arXiv:gr-qc/0604044].
\\
E.~R.~Livine and S.~Speziale,
  ``Group integral techniques for the spinfoam graviton propagator,'' to appear on JHEP
  [arXiv:gr-qc/0608131].

\bibitem{Io}
  S.~Speziale,
  ``Towards the graviton from spinfoams: The 3d toy model,''
  JHEP {\bf 05} (2006) 039 [arXiv:gr-qc/0512102].

\bibitem{3dcorr}
E. Livine, S. Speziale and J. Willis
  ``Towards the graviton from spinfoams: higher order corrections in the 3d toy model,''
arXiv:gr-qc/0605123.

\bibitem{LivineTerno}
  E.~R.~Livine and D.~R.~Terno,
   ``Reconstructing quantum geometry from quantum information: Area
  renormalisation, coarse-graining and entanglement on spin networks,''
  arXiv:gr-qc/0603008.

\bibitem{EteraPRL}
  L.~Freidel and E.~R.~Livine,
  ``3d Quantum Gravity And Effective Noncommutative Quantum Field Theory,''
  Phys.\ Rev.\ Lett.\  {\bf 96} (2006) 221301.

\bibitem{Baratin}
  A.~Baratin and L.~Freidel,
  ``Hidden quantum gravity in 3d Feynman diagrams,''
  arXiv:gr-qc/0604016.

\bibitem{Laurent3}
  L.~Freidel and D.~Louapre,
  ``Ponzano-Regge model revisited. I: Gauge fixing, observables and
  interacting spinning particles,''
  Class.\ Quant.\ Grav.\  {\bf 21} (2004) 5685
  [arXiv:hep-th/0401076].

\bibitem{asympt}
J.D.Roberts.
``Classical 6j-symbols and the tetrahedron.''
         Geometry and Topology 3 (1999), 21--66.
         [math-ph/9812013]

\bibitem{asympt2}
L.~Freidel and D.~Louapre,
``Asymptotics of 6j and 10j symbols,''
Class.\ Quant.\ Grav.\  {\bf 20} (2003) 1267
[arXiv:hep-th/0209134].
\\
J.~W.~Barrett and C.~M.~Steele,
 ``Asymptotics of Relativistic Spin Networks", Class.Quant.Grav. {\bf 20} (2003) 1341-1362, [arXiv:gr-qc/0209023].

\bibitem{Brunn}
  J.~Brunnemann and T.~Thiemann,
   ``Simplification of the spectral analysis of the volume operator in loop
  quantum gravity,''
  Class.\ Quant.\ Grav.\  {\bf 23} (2006) 1289
  [arXiv:gr-qc/0405060].

\bibitem{semi}
  C.~Rovelli and S.~Speziale,
  ``A semiclassical tetrahedron,'' Class.\ Quant.\ Grav.\ {\bf 23} (2006) 5861 - 5870,
  [arXiv:gr-qc/0606074].

\bibitem{AshtekarMeasure}
  A.~Ashtekar and J.~Lewandowski,
  ``Projective Techniques And Functional Integration For Gauge Theories,''
  J.\ Math.\ Phys.\  {\bf 36} (1995) 2170
  [arXiv:gr-qc/9411046].

\bibitem{Baez:1994hx}
  J.~C.~Baez,
  ``Spin network states in gauge theory,''
  Adv.\ Math.\  {\bf 117}, 253 (1996)
  [arXiv:gr-qc/9411007].

\bibitem{heron}
Heron of Alexandria, \emph{Metrica}, (I cen. AD).

\bibitem{Yutsis}
  D.~A.~Varshalovich, A.~N.~Moskalev and V.~K.~Khersonsky,
  \emph{Quantum Theory of Angular Momentum} (World Scientific, Singapore 1988).

\bibitem{Bar-Natan}
D. Bar-Natan, ``On the Vassiliev Knot Invariants,'' Topology {\bf 34} (1995) 423

\bibitem{DePietriScaling}
  R.~De Pietri,
   ``On the relation between the connection and the loop representation of
  quantum gravity,''
  Class.\ Quant.\ Grav.\  {\bf 14} (1997) 53
  [arXiv:gr-qc/9605064].


\end{thebibliography}
\end{document}